\documentclass[12pt]{article}
\usepackage{amsfonts,dsfont}
\usepackage{amsmath,amssymb,amsthm}
\usepackage[pdftex]{graphicx}
\usepackage{float}

\newtheorem*{thm}{Theorem}
\begin{document}

\title{Families of two-dimensional Coulomb gases on an ellipse: correlation functions and universality
}

\author{Taro Nagao$^1$, Gernot Akemann$^2$,  Mario Kieburg$^{2,3}$ and Iv\'an Parra$^2$} 

\date{}

\maketitle 

\begin{center} 
\it $^1$Graduate School of Mathematics, Nagoya University, Chikusa-ku,
\\  Nagoya 464-8602, Japan \\
\bigskip 
\it $^2$Faculty of Physics, Bielefeld University, \\ 
D-33501 Bielefeld, Germany \\
\bigskip 
\it $^3$School of Mathematics and Statistics, \\
The University of Melbourne, \\
Parkville, VIC 3010, Australia
\end{center}

\begin{abstract}
We investigate a one-parameter family of Coulomb gases in two dimensions, which are confined to an ellipse due to a hard wall constraint, 
and are subject to an additional external potential. 
At inverse temperature $\beta=2$ we can use the technique of planar orthogonal polynomials, borrowed from random matrix theory, to explicitly determine all $k$-point correlation functions for a fixed number of particles $N$. These are given by the determinant of the kernel of the corresponding orthogonal polynomials, 
which in our case are the Gegenbauer polynomials, or a subset of the asymmetric Jacobi polynomials, depending on the choice of external potential, as shown in a companion paper recently published by three of the authors. In the rotationally invariant case, when the ellipse becomes the unit disc, our findings agree with that of the ensemble of truncated unitary random matrices.
The thermodynamical large-$N$ limit is investigated in the local scaling regime in the bulk and at the edge of the spectrum at weak and strong non-Hermiticity. 
We find new universality classes in these limits and recover the sine- and Bessel-kernel in the Hermitian limit. The limiting global correlation functions of particles in the interior of the ellipse are more difficult to obtain but found in the special cases corresponding to the Chebyshev polynomials.

\end{abstract}

\medskip


\newpage
\section{Introduction}\label{intro}
\setcounter{equation}{0}
\renewcommand{\theequation}{1.\arabic{equation}}

Coulomb gases in two dimensions are constituted by a set of particles that interact logarithmically and that are 
subject to some confining potential, that may for example be given by a Gaussian or a hard wall constraint 
on a certain domain. At specific values of the temperature $T = 1/(k_B \beta)$ (with the Boltzmann constant $k_B$) they can be studied using non-Hermitian random matrix theory (RMT), where the complex matrix eigenvalues represent the locations of the charged particles. The three classical Ginibre ensembles \cite{Ginibre} for instance, which all have a Gaussian potential, correspond to one-component plasmas with a suitable background charge, 
cf. \cite{PeterCoulomb,Peterbook}.

On the one hand, Coulomb gases at general temperature $\beta$ are objects of intense study 
and pose challenging open problems, e.g. the formation of the so-called Abrikosov lattice at large $\beta$, 
and we refer to \cite{Serfatyreview} for a review. Typically, for large systems of $N\gg1$ particles with $\beta\sim\mathcal{O}(1)$, the eigenvalues condense into a droplet, the circular law for the rotationally invariant Gaussian potential, and local fluctuations around this density as well as higher order correlation functions are of interest. The case of a growing droplet where particles are constantly fed in has applications to viscous fluids, or more generally can be viewed as Laplacian growth models \cite{ZabrodinOUP}.
The case of a hard wall imposed at the edge of the droplet has been studied in 
\cite{Yacin}.
When forcing the gas away from its equilibrium position, phase transitions may occur, see \cite{CFMV} for more general potentials and the general situation in $d$ dimensions. Likewise, when $\beta= 2c/N\to0$ at fixed $c$, a smooth transition to a Gaussian is observed \cite{ABy}, including the weakly attractive case $c\in(-2,0]$. 

On the other hand, the specific value of $\beta=2$ that is tractable via RMT enjoys an exact analytical solution for finite $N$. 
Moreover, these examples find themselves in various applications e.g. in scattering in open quantum systems or in quantum field theories with chemical potential, cf. \cite{YHJS} and \cite{QCD} for respective reviews.
A powerful technique providing an exact solution of such Coulomb gases uses orthogonal polynomials in the complex plane. Exploiting the fact that at $\beta=2$ the joint density of complex eigenvalues forms a determinantal point process, one can explicitly construct the kernel of such planar polynomials and thus determine all eigenvalue correlation functions. 
Taking the complex elliptic Ginibre ensemble as an example, which is not rotationally invariant and supported on 
the full complex plane $\mathbb{C}$, these planar polynomials are provided by the Hermite polynomials \cite{FKSlong}. 
They are orthogonal with respect to a Gaussian weight function, with different variances in real and imaginary parts \cite{vEM,PdF}.
Based on the explicit solution for the kernel various large-$N$ limits can be taken. At strong non-Hermiticity, the global eigenvalue density condenses onto an ellipse in the complex plane. Nevertheless, the local eigenvalue correlation functions at the edge and in the bulk of the spectrum agree with that of the rotationally invariant complex Ginibre ensemble. In fact much further reaching universality results for complex Wigner ensembles are known \cite{TaoVu}.

A particularly interesting limit called weak non-Hermiticity was introduced in \cite{FKSlett} for the elliptic Ginibre ensemble. Whereas in this limit the global density collapses to the semi-circle on the real line, locally correlations of $\mathcal{O}(1/N)$ still extend into the complex plane. In the bulk the limiting kernel at weak non-Hermiticity is a one-parameter deformation of the celebrated sine-kernel, known from one-dimensional Wigner-Dyson statistics in RMT, which is highly universal \cite{ArnoOUP}. 
The universality of this deformed, weakly non-Hermitian kernel was first shown heuristically in \cite{FKSlong}, using supersymmetry for independent matrix elements, and more recently proven for a class of non-Gaussian deformations \cite{ACV}, including fixed trace ensembles which are non-determinantal.
This concept of weak non-Hermiticity was applied to other ensembles \cite{ZS,Osborn} and different scaling limits were found also at hard \cite{ZS,Osborn} and soft edges \cite{Bender} of the spectrum, cf. \cite{APh} for a list of many known kernels that deform the three classical ensembles and their chiral counterparts. For the scaling limit in the vicinity of a cusp or close to a hard wall we refer to \cite{Yacin,Yacin2}.

In this paper we will take the large-$N$ limit of a new class of Coulomb gases that are confined by a hard wall constraint to live on an ellipse at finite-$N$ already, subject to an additional potential.
The solution is based on another class of classical orthogonal polynomials that were shown in a companion 
paper \cite{ANPV} to be orthogonal on such a domain, subject to certain families of external potentials: 
the Gegenbauer (or ultraspherical) polynomials, which are the Jacobi polynomials with symmetric indices, 
and a subset of the Jacobi polynomials with unequal ones. 
At present we do not have a non-trivial random matrix representation for the determinantal point process solved by these polynomials. Only in the rotationally invariant case, when the ellipse degenerates to the unit disc, 
it follows from the complex eigenvalue distribution of truncated unitary matrices, with monomial orthogonal polynomials \cite{ZS}.

The outline of this paper is the following. 
In Section \ref{our model} we introduce the family of Coulomb gases that we will study and discuss limiting cases to known results in two and one dimensions.
Section \ref{corr} reviews the determinantal structure of these 
at the special inverse temperature $\beta=2$, cf. \cite{ANPV}. The corresponding planar orthogonal Gegenbauer polynomials and 
their corresponding kernel are presented. 
The limits to known kernels are given, in order to prepare a later comparison of the microscopic kernels.
Section \ref{sec:weaklim} comes to our new results and is devoted to the local, microscopic correlations 
in the weak non-Hermiticity limit. 
Subsection \ref{weak-bulk} deals with the scaling limit in the bulk, close to the origin, then turning to the edge scaling limit in Subsection \ref{weak-edge}. In both limits we find new one-parameter universality classes 
deforming the sine- and Bessel-kernel that we recover in the Hermitian limit. A large weak non-Hermiticity parameter 
is known to lead to strong non-Hermiticity, which we thus explore indirectly. In the bulk we find a new 
limiting kernel as well, and recover a well-known bulk result (the Ginibre kernel) in a limit of a potential parameter. 
At the edge, on the other hand, we recover the result from the truncated unitary matrix ensemble. As a further 
check the edge kernel is found to be asymptotically similar to the bulk kernel, thus underlining its conjectured universality. The global large-$N$ limit is addressed in Section \ref{weight1} for a special case of the Chebyshev polynomials of the second kind, 
being orthogonal with respect to the flat measure \cite{Henrici}. Two families of the non-symmetric Jacobi polynomials and their 
corresponding Coulomb gases that were introduced in \cite{ANPV} for finite $N$ are analysed in Appendix \ref{AJacobi}, giving rise to two further local universality 
classes at the edge. In Appendices \ref{BChebyshev} and \ref{CChebyshev}, the global regime is again considered for Coulomb gases related to the remaining Chebyshev polynomials that were known to be orthogonal, cf. \cite{MH}.

\section{A family of Coulomb gases on an ellipse}\label{our model}
\setcounter{equation}{0}
\renewcommand{\theequation}{2.\arabic{equation}}

In this section we will introduce the particular Coulomb gas that we will investigate. We also point out limits to systems of charged particles previously known from RMT.
Let us consider a two-dimensional, static one-component 
Coulomb gas  with a Hamiltonian 
\begin{equation}\label{H}
H = \sum_{j=1}^N V(z_j) - \sum_{j<l}^N \log|z_j - z_l|\ .
\end{equation}
The locations of the 
particles interacting logarithmically in the plane are
denoted by complex numbers $z_j = x_j + i y_j$ ($j = 
1,2,\cdots,N$) with the standard map $(x_j,y_j) \in \mathbb{R}^2 \mapsto z_j \in \mathbb{C}$. 
We impose the particles to be  
confined to an ellipse, which is given in the following parametrisation.
\begin{equation}
\label{ellipse}
E = \left\{z=x+iy \left| \frac{2 \tau}{1 + \tau} x^2 + \frac{2 \tau}{1 - \tau}
  y^2 \leq 1 \right. \right\}\ ,\quad 0 < \tau < 1.
\end{equation}
Here $x$ and $y$ are real. The one-particle potential in the Hamiltonian \eqref{H} is 
given by
\begin{equation}
V(z) = - \frac{a}{2} \log\left( 1 - \frac{2 \tau}{1 + \tau} x^2 - 
\frac{2 \tau}{1 - \tau} y^2 \right)\ , \quad a>-1\ .
\end{equation}
This potential mimics a charged mirror at the boundary of the ellipse which is either attractive ($a  < 0$) or 
repulsive ($a >0$). The resulting probability distribution function for the particles to be at equilibrium at an inverse temperature $1/(k_BT)= \beta = 2$ is known to be
\begin{equation}
\label{pdf}
P(z_1,z_2,\cdots,z_N) = \frac{1}{Z_N} e^{- \beta H} 
= \frac{1}{Z_N} \prod_{j=1}^N w(z_j) \prod_{j<l}^N |z_j - z_l|^2\ .
\end{equation}
Here, we define a one-particle weight function
\begin{equation}
\label{weight}
w(z) = \left(1 - \frac{2 \tau}{1 + \tau} x^2 - 
\frac{2 \tau}{1 - \tau} y^2 \right)^a = e^{-\beta V(z)},
\end{equation}
which is real and non-negative, $w(z)=w(\bar{z}) \geq 0$ ($\forall z \in E$), 
and use the integration measure $\prod_{j=1}^N d^2z_j = 
\prod_{j=1}^N dx_j dy_j$.  The notation  ${\bar z}$ denotes the complex conjugate of $z$. 
The point process in \eqref{pdf} is determinantal, as shown in Section \ref{corr}, cf.\cite{ANPV}. 
The partition function that normalises  the distribution \eqref{pdf} is defined as 
\begin{equation}
\label{ZN}
Z_N=\prod_{j=1}^N \int_E d^2z_j \ w(z_j) \prod_{i<l}^N |z_i - z_l|^2\ .
\end{equation}

Let us point out several limits of the distribution \eqref{pdf} known from RMT. 
First, we consider the rotationally invariant limit. Here, we have to rescale 
the positions as
\begin{equation}
\label{tscale}
x_j \mapsto x_j/\sqrt{2 \tau}, \ \ \ y_j \mapsto y_j/\sqrt{2 \tau}\ ,
\end{equation}
and then take the limit $\tau \rightarrow 0$. In this limit the ellipse $E$ in \eqref{ellipse} becomes the unit disc. The limiting weight function becomes 
\begin{equation}
\label{truncated}
w_{truncated}(z) = \left(1 - |z|^2 \right)^a\ ,\quad a>-1\ ,
\end{equation}
which is radially symmetric. For an integer $a$ the limiting joint density from \eqref{pdf} then agrees with the distribution of the complex eigenvalues of the ensemble of truncated 
unitary random matrices introduced in \cite{ZS}. It is obtained from a unitary matrix $U\in U(N)$ distributed according to the Haar measure, truncated to the upper left block of $U$ of size $M\times M$, with $N>M$ and the resulting parameter
\begin{equation}
\label{adef}
a=N-M-1\ .
\end{equation}
The complex eigenvalue correlation functions of such a truncated unitary matrix were computed in \cite{ZS}, using monomials $M_n(z)=z^n$ as orthogonal polynomials with respect to the weight \eqref{truncated}.

In the second limit, we want to make contact with the eigenvalues of Hermitian RMT and 
the corresponding Dyson gas of particles confined to (a subset of) the real line, while still interacting 
logarithmically, that is with  Coulomb interaction in two dimensions.  
Taking the limit $\tau \rightarrow 1$ on the ellipse $E$ in \eqref{ellipse} enforces the imaginary part 
to condense on a narrow strip about the real line and eventually to vanish, $y\to0$, and thus maps $E$ to the interval $[-1,1]$. Because the initial measure is in two dimensions, 
in \eqref{pdf} we still have to integrate out the imaginary parts $\Im (z_j)=y_j$, leading to an additional contribution to the weight function, see \cite[Remark 3.7]{ANPV} for details. We arrive at the following limiting weight function 
\begin{equation}\label{Jacobi}
w_{Jacobi}(x) = \left(1 - x^2 \right)^{a+\frac12},
\end{equation}
with joint density \eqref{pdf} projected to the real parts $\Re (z_j)=x_j\in[-1,1]$ ($j= 1,2,\cdots,N$).
It agrees with a special case of the weight where the eigenvalues result from the Jacobi ensemble of Hermitian 
random matrices \cite{FK,NW}. The eigenvalue correlation functions are computed with the help of 
the Jacobi polynomials, in our case with symmetric indices, when the Jacobi polynomials reduce to the 
Gegenbauer polynomials (also called the ultraspherical polynomials). At $a=0$ these become the Chebyshev polynomials of the second kind.

Finally, as also pointed out in \cite{ANPV}, a map to the elliptic Ginibre ensemble exists,  
thus removing the hard wall constraint, with $E$ becoming the entire complex plane after rescaling. This is 
achieved by making the scaling transformations (for $a>0$) 
\begin{equation}\label{ascale}
x_j \mapsto x_j/\sqrt{2\tau a}, \ \ \ y_j \mapsto y_j/\sqrt{2\tau a}\ , 
\end{equation}
and then taking the limit $a \rightarrow \infty$. Hence, the particles are pushed away from the 
boundary until it has no contact at all. Due to the scaling we zoom into the origin and find 
the limiting weight function \eqref{weight} which is a Gaussian, 
\begin{equation}
\label{Gauss}
w_{Ginibre}(z) = {\rm exp}\left(- \frac{1}{1 + \tau} x^2 - 
\frac{1}{1 - \tau} y^2 \right).
\end{equation}
The resulting limiting distribution \eqref{pdf} agrees with that of the complex eigenvalues
of the elliptic Ginibre ensemble of complex random matrices \cite{SCSS}, including the rotationally invariant Ginibre ensemble at $\tau=0$.
The elliptic Ginibre ensemble was analysed as a Coulomb gas in \cite{PdF}, deriving and using the 
orthogonality property of the Hermite polynomials with respect to the weight \eqref{Gauss}. 
All complex eigenvalue correlation functions of the elliptic Ginibre ensemble were derived later in \cite{FKSlong}.

The rotationally invariant limit \eqref{truncated} and the real limit \eqref{Jacobi} 
will provide us with consistency checks for our Coulomb gas in the large-$N$ limit,
and lead to a better understanding of the issue of universality. A comparison to the elliptic Ginibre ensemble 
is more difficult which is related to the fact that its initial support is the full complex plane. Even after 
taking the large-$N$ limit, the correlations at the edge of the limiting elliptic support only decay exponentially, 
in contrast to the hard constraint present in our case. This difference will be discussed in more detail 
in Subsection \ref{weak-bulk}.


\section{Density correlation functions at finite-$N$}
\label{corr}
\setcounter{equation}{0}
\renewcommand{\theequation}{3.\arabic{equation}}

Let us recall that the probability distribution functions of the form (\ref{pdf}) with $\beta=2$ is a 
determinantal point process. Thus all density correlation functions are given in terms of a kernel of orthogonal polynomials in the complex plane. Suppose that 
the polynomials 
$M_n(z) = z^n +\mathcal{O}(z^{n-1})$ 
in monic normalisation satisfy the following orthogonality relation 
\begin{equation}
\label{OPM}
\int_D d^2z \ w(z) M_m(z) M_n({\bar z}) =  h_n \delta_{m,n}\ ,\ \ m,n = 0,1,2,\cdots,
\end{equation}
for a given non-negative weight function $w$ on some domain $D$. Here, $z = x + i y$ ($x$ and $y$ are real) 
and $d^2z = dx dy$. In our case the integration 
domain $D$ is given by the ellipse $E$ in (\ref{ellipse}), and 
the weight function $w(z)$ (satisfying $w(z) = w({\bar z})$) is (\ref{weight}). 
Then, in general  the $k$-point density correlation function defined as 
\begin{equation}
\rho(z_1,z_2,\cdots,z_k) = \frac{N!}{(N-k)!} \int_D d^2z_{k+1} \int_D
d^2z_{k + 2} \cdots \int_D d^2z_N P(z_1,z_2,\cdots,z_N), 
\end{equation}
can be written in a determinantal form, in terms of the kernel $K_{N}$ of these polynomials $M_n(z)$ from \eqref{OPM}, cf. \cite{KS}:
\begin{equation}
\label{det}
\rho(z_1,z_2,\cdots,z_k) = {\rm det}\left[ K_N(z_j, z_l) \right]_{j,l = 1,2,\cdots,k} .
\end{equation}
Here, the kernel is given by the sum over the orthonormalised polynomials, 
\begin{equation}
\label{detkernel}
K_N(z_j,z_l) = \sqrt{w(z_j) w({\bar z_l})} \sum_{n = 0}^{N-1} \frac{1}{h_n} M_n(z_j) M_n({\bar z_l}).
\end{equation}
Therefore, in order to see the asymptotic behaviour of the correlation 
functions in any particular limit $N \rightarrow \infty$, we only need to evaluate 
the limit of the kernel  $K_N(z_1,z_2)$.

Let us now specify the polynomials for our elliptic domain \eqref{ellipse}. In \cite{ANPV} the following 
orthogonality relation was proven\footnote{Compared to \cite{ANPV} where the ellipse is parametrised by 
$(x/a)^2+(y/b)^2\leq1$ with $a>b>0$, we have chosen a one-parameter family, setting $a^2=(1+\tau)/2\tau$ 
and $b^2=(1-\tau)/2\tau$ and thus $a^2-b^2=1$, with foci located at $z = \pm1$.}:
\begin{eqnarray}
\label{orthogonality}
\int_E d^2z \left( 1 - \frac{2 \tau}{1 + \tau} x^2 - \frac{2 \tau}{1 - \tau} 
y^2 \right)^a C^{(a + 1)}_m(z) C^{(a + 1)}_n({\bar z}) &\nonumber \\  =   
\frac{\sqrt{1 - \tau^2}}{2 \tau} \frac{\pi}{n + a + 1} C^{(a + 1)}_n\left( 
\frac{1}{\tau} \right) \delta_{m,n}&,\quad a>-1,
\end{eqnarray}
for $m,n = 0,1,2,\cdots$ on the ellipse $E$ in \eqref{ellipse}. The polynomials $C^{(a+1)}_n$ are  the Gegenbauer polynomials given by 
\begin{equation}
\label{gegenbauer}
C^{(a + 1)}_n(z) = \sum_{j=0}^{\lfloor n/2 \rfloor} \frac{(-1)^j \Gamma(n + a - j + 1)}{
\Gamma(a + 1) \Gamma(j+1) \Gamma(n - 2 j + 1)} (2 z)^{n - 2 j}\ ,
\end{equation}
where $\lfloor  n/2 \rfloor$ is the floor function, meaning the greatest integer that is less than 
or equal to $n/2$. Equivalently, they can be expressed in terms of Gau{\ss}' hypergeometric 
function, or in terms of the Jacobi polynomials \cite{szego}
\begin{equation}
P^{(\alpha,\gamma)}_n(z) =  \frac{1}{(1 - z)^\alpha (1 + z)^\gamma} \frac{(-1)^n}{2^n n!} 
\frac{d^n}{dz^n} \left[ (1 - z)^{n + \alpha} (1 + z)^{n + \gamma} \right],
\end{equation}
with symmetric indices $\displaystyle \alpha = \gamma = a + \frac{1}{2}$, cf. \eqref{CP} below. In particular 
they have parity symmetry, 
$C^{(a+1)}_n(-z)=(-1)^nC^{(a+1)}_n(z)$.
We find that the corresponding monic orthogonal polynomials read 
\begin{equation}
\label{MCrel}
M_n(z) = \frac{\Gamma(a + 1) \Gamma(n + 1)}{\Gamma(n + a + 1) 2^n} 
C^{(a + 1)}_n(z)\ ,
\end{equation}
and that the normalisation constants resulting from \eqref{orthogonality} are obtained as 
\begin{equation}
\label{norms}
h_n = \frac{\Gamma(a + 1) \Gamma(n + 1)}{\Gamma(n + a + 2) 2^n} 
\frac{\pi\sqrt{1 - \tau^2}}{2 \tau} 
M_n\left(\frac{1}{\tau}\right).
\end{equation}
Notice that for $n=0$ we obtain the normalisation of our weight over $E$,
\begin{equation}
\label{Dnorm}
A=\int_E d^2z \left( 1 - \frac{2 \tau}{1 + \tau} x^2 - \frac{2 \tau}{1 - \tau} 
y^2 \right)^a = \frac{1}{a + 1} \frac{\pi\sqrt{1 - \tau^2}}{2 \tau} \ .
\end{equation}
In \cite{ANPV} the orthogonality of two families of the Jacobi polynomials with non-symmetric 
indices on an ellipse was also derived. Their analysis is deferred to Appendix \ref{AJacobi}.

In the special case $a=0$, when the Gegenbauer polynomials reduce to the Chebyshev polynomials of the second kind, $U_n(z)=C^{(1)}_n(z)$, the proof of the orthogonality relation (\ref{orthogonality}) was previously known, see
\cite{Henrici,MH}. 
It follows that the kernel $K_N(z_1,z_2)$ of the orthonormalised Gegenbauer polynomials is given by
\begin{eqnarray}
\label{kernel}
K_N(z_1,z_2) & = & \left(1 - \frac{2 \tau}{1 + \tau} x_1^2 - \frac{2 \tau}{1 - \tau} 
y_1^2 \right)^{a/2} \left(1 - \frac{2 \tau}{1 + \tau} x_2^2 - \frac{2 \tau}{1
  - \tau} y_2^2 \right)^{a/2} \nonumber \\ 
&& \times  \frac{2 \tau}{\pi \sqrt{1 - \tau^2}} \sum_{n=0}^{N-1} 
\frac{n + a + 1}{C^{(a + 1)}_n(1/\tau)} C^{(a + 1)}_n(z_1) C^{(a + 1)}_n({\bar z_2})\ .
\end{eqnarray}
This completes the computation of all correlation functions via \eqref{det} for finite-$N$.

Before  evaluating the asymptotic  of this kernel in various limits $N \rightarrow 
\infty$ in the following sections, let us show how the kernel reduces to known limiting cases at finite-$N$, 
the rotationally invariant case and the Hermitian limit. 

\begin{figure}[H]
\begin{center}
\includegraphics[height=6cm]{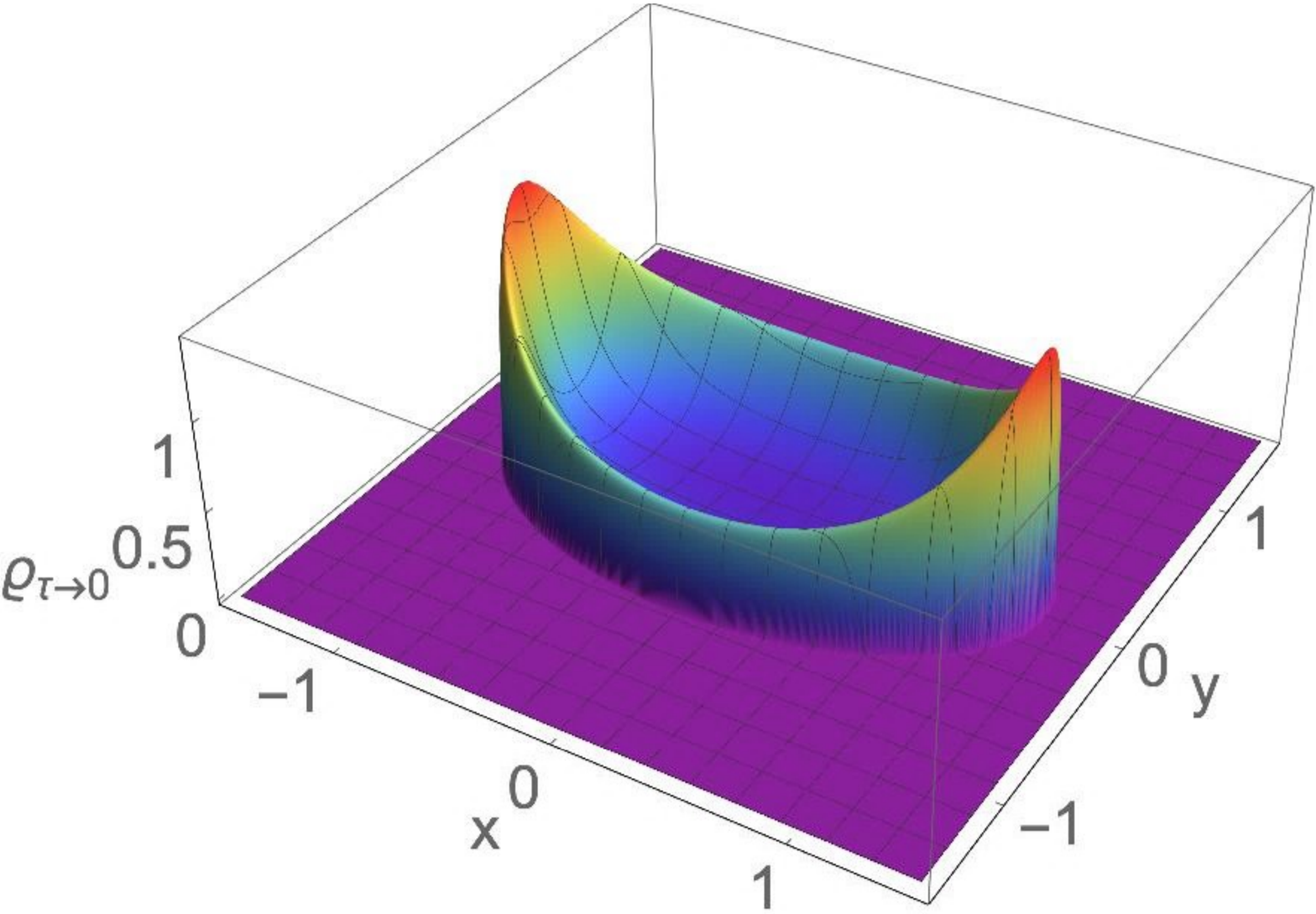}
\includegraphics[height=6cm]{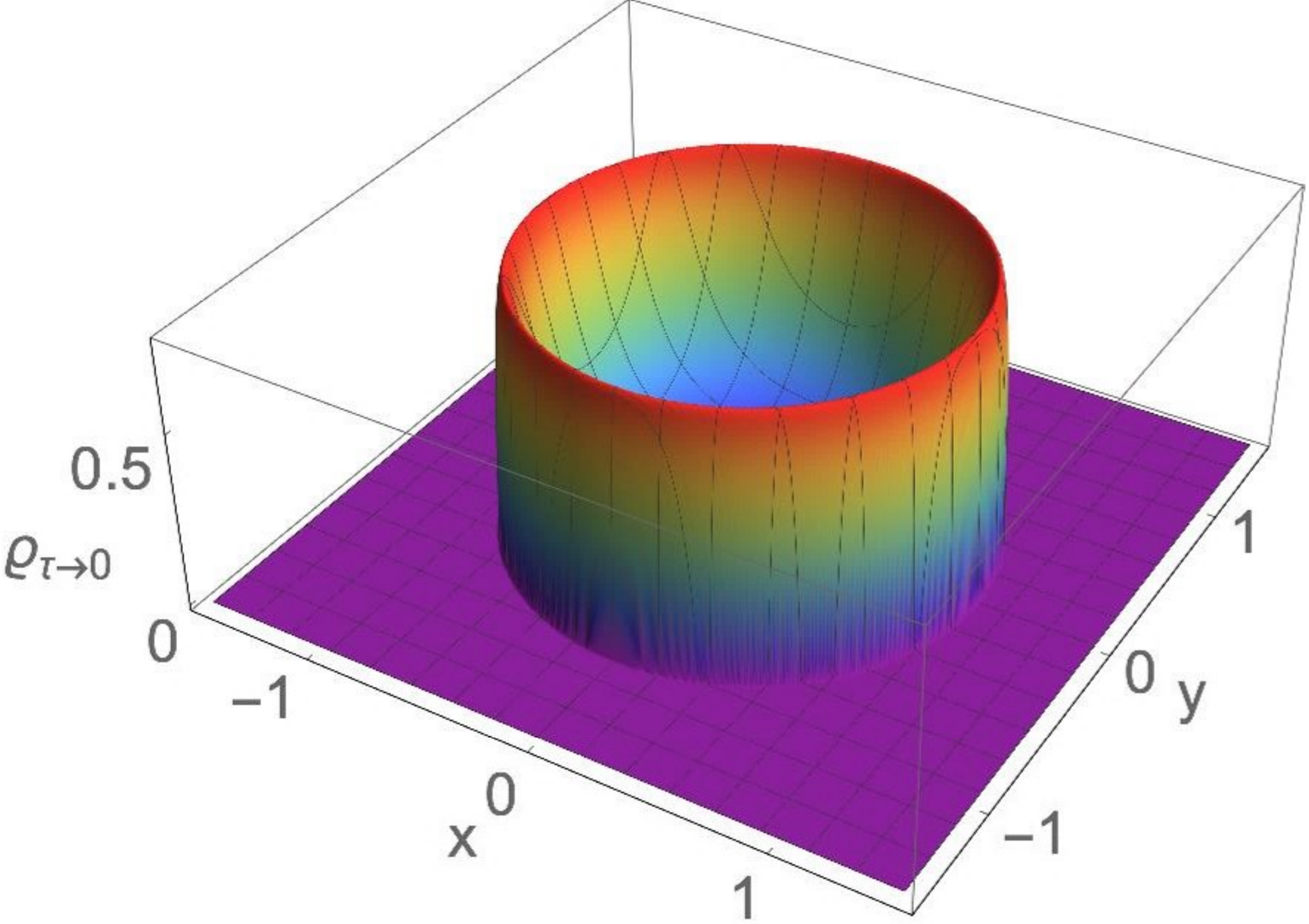}
\caption{\small 
The rescaled particle densities
$
\rho_{\tau \to 0}=\frac{1}{2\tau N} 
K_N\left(\frac{x+iy}{\sqrt{2\tau}},\frac{x+iy}{\sqrt{2\tau}} \right)
$
for $N=10$, $a=1$, and $\tau=0.5$ (upper figure) as well as $\tau=0.005$ (lower figure).}
\label{fig.01}
\end{center}
\end{figure}

We begin with the rotationally invariant limit. After the rescaling \eqref{tscale} and sending $\tau\to0$, we map 
the ellipse \eqref{ellipse} to the unit disc, $E\to\{z = x + i y\,|\,x^2+y^2\leq1\}$. The orthogonal 
polynomials  of the limiting weight function 
(\ref{truncated}) are now monomials, $M_n(z) = z^n$, and the orthogonality relation is given by
\begin{equation}
\int_{|z| \leq 1} d^2z (1 - |z|^2)^a z^m {\bar z}^n = h_n^{truncated} \delta_{m,n},
\end{equation}
with norms
\begin{equation}
h_n^{truncated} = \pi \frac{\Gamma(a + 1) \Gamma(n + 1)}{\Gamma(n + a + 2)}\ .
\end{equation}
For the limit of the kernel \eqref{kernel} we obtain
\begin{eqnarray}
\label{rs}
K_N^{truncated}(z_1,z_2)&=&
\lim_{\tau\to0}\frac{1}{2\tau}
K_N\left(\frac{z_1}{\sqrt{2\tau}},\frac{z_2}{\sqrt{2\tau}}\right) 
\nonumber \\ 
&=&
(1 - |z_1|^2)^{\frac{a}{2}} (1 - |z_2|^2)^{\frac{a}{2}} \sum_{n=0}^{N-1} 
\frac{\Gamma(n + a + 2)}{\pi \Gamma(a + 1) \Gamma(n + 1)} 
(z_1 {\bar z}_2)^n.
\end{eqnarray}
For non-negative integer values of $a$ 
it agrees with the kernel derived in \cite{ZS} for the ensemble of truncated unitary random 
matrices, with relation \eqref{adef} between the parameter $a$ and the matrix dimensions. 
The rescaled $1$-point density correlation functions (particle densities) describing the 
approach to the rotationally invariant limit $\tau \to 0$ are illustrated in Fig. \ref{fig.01}. 
As the plot clearly highlights the spectrum concentrates on the one-dimensional ellipse and has 
only exponential tails into its interior.

\begin{figure}[H]
\begin{center}
\includegraphics[height=6cm]{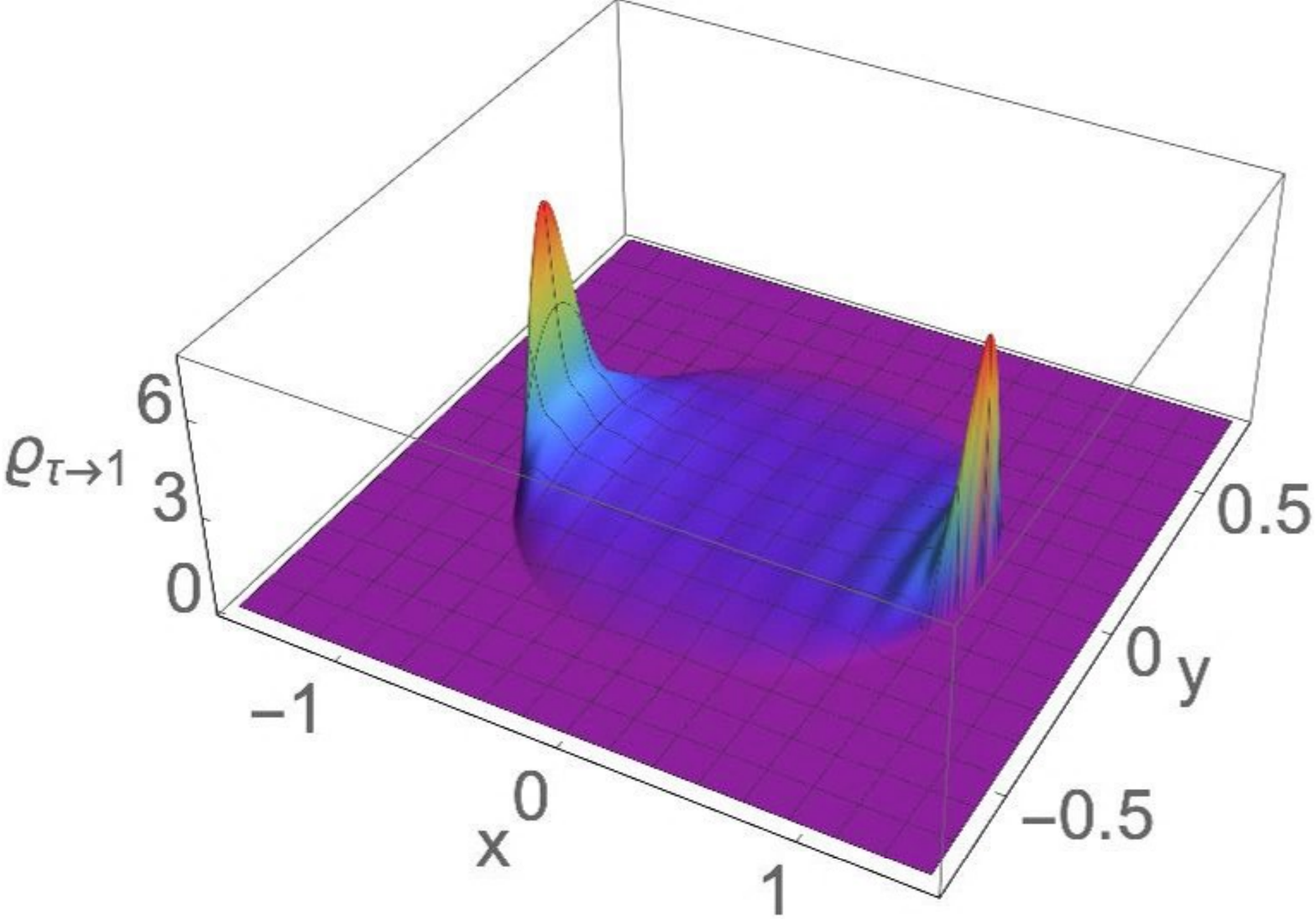}
\includegraphics[height=6cm]{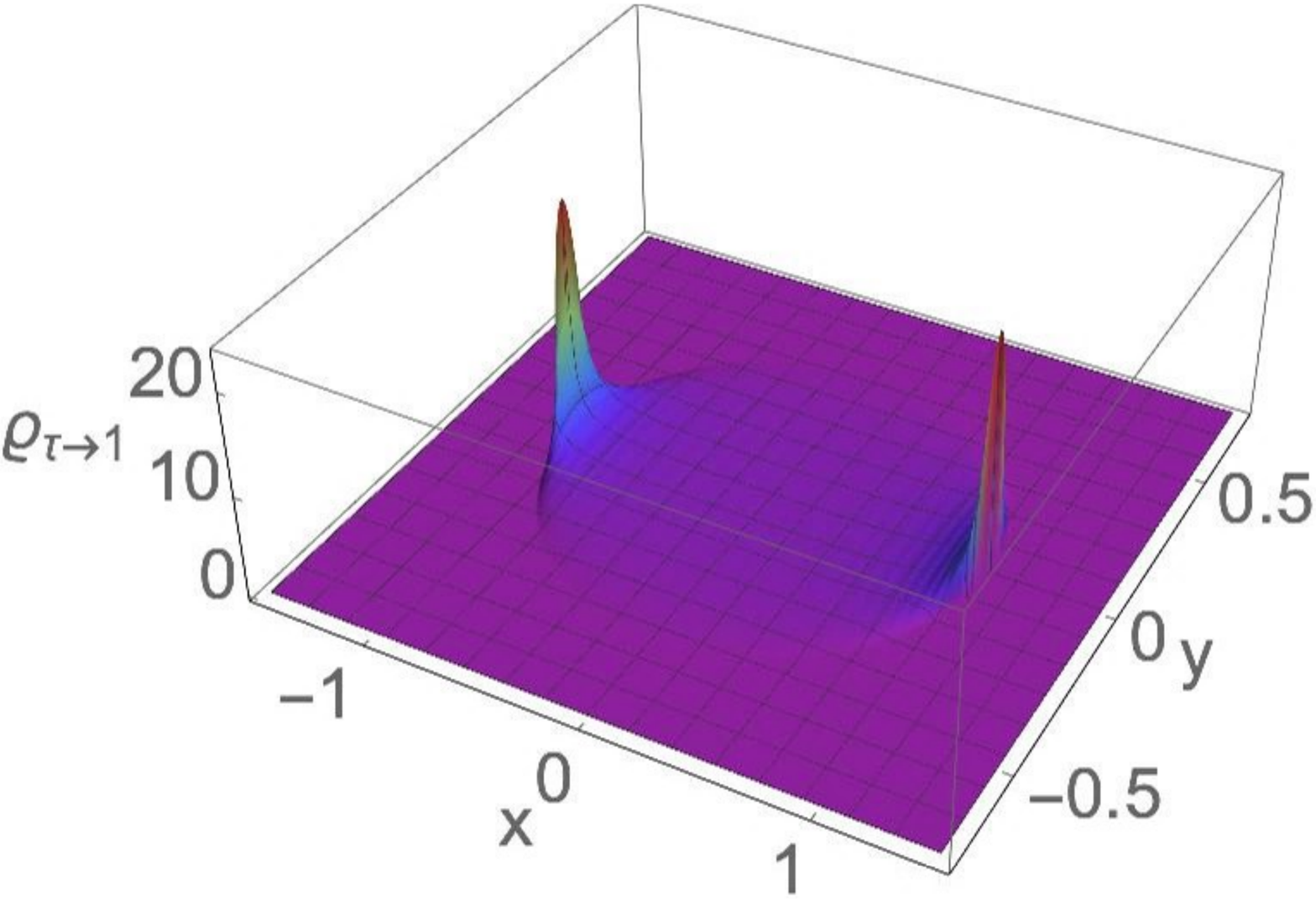}
\caption{\small 
The rescaled particle densities
$
\rho_{\tau \to 1}=\frac{1}{N^2} 
K_N\left( x+i\frac{y}{N},x+i\frac{y}{N}\right)
$
for $a=1$, $\tau=\left(1+\frac{s^2}{2N^2}\right)^{-1}$ with $s=1$, and $N=10$ (upper figure) as well as $N=30$ (lower figure).}
\label{fig.02}
\end{center}
\end{figure}

In the Hermitian limit $\tau\to1$ the ellipse \eqref{ellipse} is mapped to $[-1,1]$. In order to be able to take the limit of the orthogonality relation \eqref{OPM}, with \eqref{MCrel} and \eqref{norms}, 
we have to divide the orthogonality relation by the normalisation $A$ from \eqref{Dnorm}, yielding the 
normalised integral, cf. \cite{ANPV}
\begin{equation}
\label{DIrel}
1=\lim_{\tau\to1} \frac{1}{A}\int_E d^2z \left( 1 - \frac{2 \tau}{1 + \tau} x^2 - \frac{2 \tau}{1 - \tau} 
y^2 \right)^a =\frac{1}{B} \int_{-1}^1 dx (1 - x^2)^{a+\frac12} ,
\end{equation}
with 
\begin{equation}
\label{Bdef}
B=\frac{\sqrt{\pi}\Gamma\left(a+\frac32\right)}{\Gamma(a+2)}\ .
\end{equation}
The limit $\tau\to1$ of the monic polynomials $M_n(z)$ is non-singular (and remains monic), and due to the relation 
between the Gegenbauer and Jacobi polynomials with symmetric indices \cite{NIST},
\begin{equation}
\label{CP}
C_{n}^{(a+1)}(z)
=\frac{\displaystyle \Gamma(n+2a+2)  \Gamma\left(a+\frac32\right)}{\displaystyle \Gamma(2 a + 2) 
\Gamma\left(n + a+\frac32\right)} P_n^{(a+\frac12, \ a+\frac12)}(z)\ ,
\end{equation}
they can also be expressed in terms of the latter,
\begin{equation}
\label{MPrel}
M_n(z)= \frac{2^n \Gamma(n+1) \Gamma(n+2a+2)}{\Gamma(2n+2a+2)} 
P_n^{(a+\frac12, \ a+\frac12)}(z)\ .
\end{equation}
It remains to evaluate $M_n(1)$ in the limiting norms \eqref{norms}, where we can use \cite{NIST}
\begin{equation}
\label{C1}
C_n^{(a+1)}(1)=\frac{\Gamma(n+2a+2)}{\Gamma(2a+2)\Gamma(n+1)},
\end{equation}
together with \eqref{MCrel}. Inserting all ingredients, using the doubling formula 
for the Gamma function
\begin{equation}
\label{double}
\sqrt{\pi}\Gamma(2z)=2^{2z-1}\Gamma(z)\Gamma\left(z+\frac12\right),
\end{equation}
and multiplying with $B$ after taking the limit \eqref{DIrel}, we arrive at the following orthogonality 
relation for the weight \eqref{Jacobi}  
\begin{equation}
\int_{-1}^1 dx (1 - x^2)^{a+\frac12} M_m(x) M_n(x) = h_n^{Jacobi} \delta_{m,n},
\end{equation}
with 
\begin{equation}
\label{Jacobinorm}
h_n^{Jacobi}= \frac{\pi\Gamma(n+1)\Gamma(n+2a+2)}{2^{2n+2a+1}\Gamma(n+a+2)\Gamma(n+a+1)}
\ .
\end{equation}
Together with \eqref{MPrel} this agrees with the standard orthogonality relation of the Jacobi 
polynomials \cite{Peterbook}. We have drawn the rescaled 1-point density correlation functions 
(particle densities) for this limit in Fig. \ref{fig.02}. The spectrum evidently concentrates on the two 
extremal points $z=\pm1$ because the Hermitian Jacobi ensemble, to which it converges, has square 
root singularities at these points.

\begin{figure}[H]
\begin{center}
\includegraphics[height=6cm]{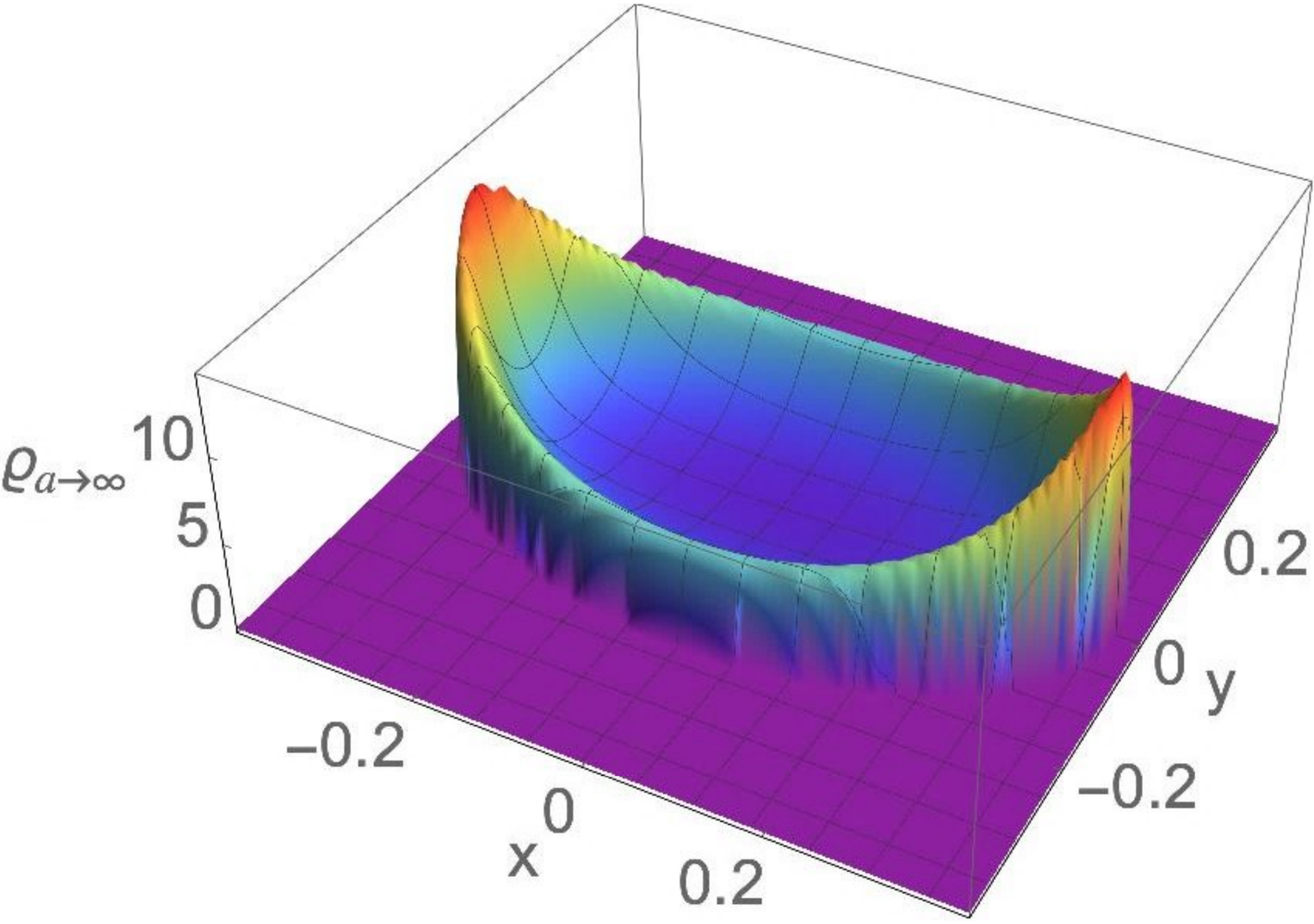}
\includegraphics[height=6cm]{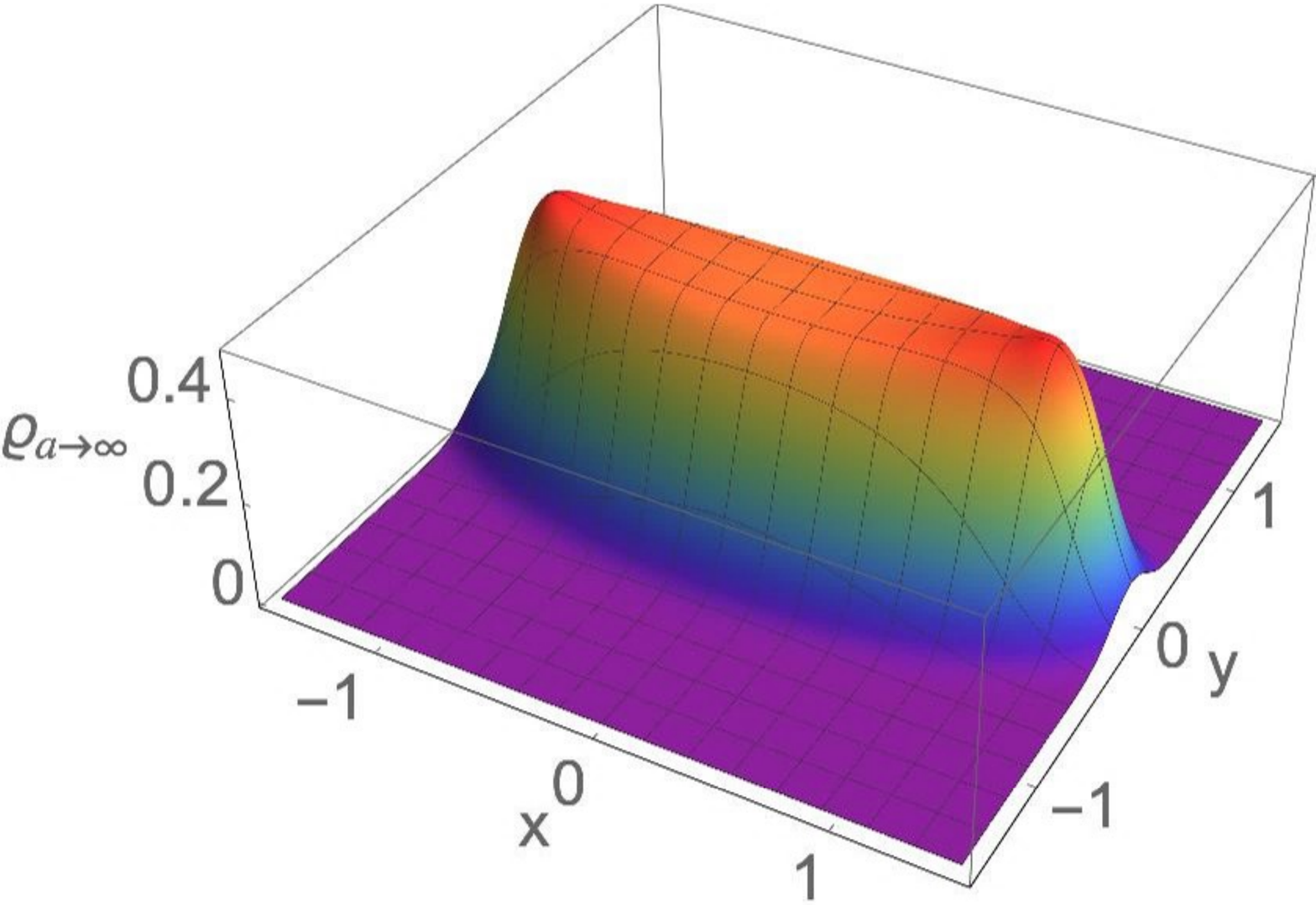}
\caption{\small 
The rescaled particle densities
$
\rho_{a\to\infty}=\frac{1}{2\tau a}
K_N\left(\frac{\sqrt{N}(x+iy)}{\sqrt{2\tau
a}},\frac{\sqrt{N}(x+iy)}{\sqrt{2\tau a}}\right)
$
for $N=10$, $\tau=0.5$, and $a=1$ (upper figure) as well as $a=100$ (lower figure).}
\label{fig.03}
\end{center}
\end{figure}

It is also possible to recover the Hermite polynomials $H_n(z)$, which are orthogonal 
with respect to the weight \eqref{Gauss} in the full complex plane \cite{PdF}, after 
taking the scaling limit $a\to\infty$ with \eqref{ascale}. This limit to the elliptic Ginibre 
ensemble requires more preparation. While 
the limit of Gegenbauer polynomials with rescaled argument, as required 
by \eqref{ascale}, is well-known \cite[18.7.24]{NIST}:
\begin{equation}\label{Clim1}
\lim_{a\to\infty}a^{-n/2}C_n^{(a)}(z/\sqrt{a})= \frac{1}{n!} H_n(z) \ ,
\end{equation}
for the norm \eqref{norms} we also need the corresponding limit {\it 
without} rescaling the arguments. It follows from the generating 
function for Gegenbauer polynomials \cite[18.12.4]{NIST}
\begin{equation}\label{Cgen}
\sum_{n=0}^\infty C_n^{(a)}(x)r^n=(1-2rx+r^2)^{-a}\ .
\end{equation}
After rescaling $r\to r/a$ and taking $a\to\infty$,
\begin{eqnarray}
\lim_{a\to\infty}\sum_{n=0}^\infty \frac{1}{a^n} C_n^{(a)}(x)r^n&=& 
e^{2rx}= \sum_{n=0}^\infty \frac{(2x)^n}{n!} r^n\ ,
\end{eqnarray}
we obtain the relation
\begin{equation}\label{Clim2}
\lim_{a\to\infty} a^{-n} C_n^{(a)}(x)=\frac{1}{n!} (2x)^n\ .
\end{equation}
Note the difference in the power of $a$ compared to \eqref{Clim1}. 
Putting these together and rescaling as in \eqref{ascale}, we obtain
\begin{eqnarray}
\label{kernelGin}
K_N^{Ginibre}(z_1,{z}_2) & = & \lim_{a\to\infty}\frac{1}{2\tau a}K_N\left(\frac{z_1}{\sqrt{2\tau 
a}},\frac{z_2}{\sqrt{2\tau a}}\right) \nonumber \\ 
&=&\exp\left[-\frac{x_1^2}{2(1+\tau)}-\frac{y_1^2}{2(1-\tau)}
-\frac{x_2^2}{2(1+\tau)}-\frac{y_2^2}{2(1-\tau)}\right]\nonumber\\
&&\times \frac{1}{\pi \sqrt{1 - \tau^2}} \sum_{n=0}^{N-1} 
\left(\frac{\tau}{2}\right)^n\frac{1}{n!}
H_n\left( \frac{z_1}{\sqrt{2\tau}}\right)H_n\left( 
\frac{\bar{z}_2}{\sqrt{2\tau}}\right).
\end{eqnarray}
It agrees with the kernel of the elliptic Ginibre ensemble 
\cite{FKSlong}. The Hermite polynomials satisfy \cite{vEM,PdF}
\begin{eqnarray}\label{HOP}
\int_{\mathbb{C}}d^2z 
\exp\left[-\frac{x^2}{1+\tau}-\frac{y^2}{1-\tau}\right] H_n\left( 
\frac{z}{\sqrt{2\tau}}\right)H_m\left( 
\frac{\bar{z}}{\sqrt{2\tau}}\right)
&=&h_n^{Ginibre}\delta_{n,m}\ ,\\
h_n^{Ginibre}&=& n!{\pi \sqrt{1 - \tau^2}} 
\left(\frac{\tau}{2}\right)^{-n}\ ,\nonumber
\end{eqnarray}
for $0 < \tau < 1$. As before, we illustrated the rescaled 1-point density correlation functions 
(particle densities) in the limit to the elliptic Ginibre ensemble $a\to\infty$, see Fig. \ref{fig.03}. 
This time the spectrum fills out the whole ellipse so that a well defined bulk is available.

\section{Local correlations at weak non-Hermiticity}
\label{sec:weaklim}
\setcounter{equation}{0}
\setcounter{figure}{0}
\renewcommand{\theequation}{4.\arabic{equation}}

In this section we come to our new results and will mainly be concerned with local correlation functions in the weakly non-Hermitian situation. 
For a discussion of strong non-Hermiticity we refer to the respective Subsections \ref{weak-bulk} and \ref{weak-edge}.
With weak non-Hermiticity we mean 
a double scaling limit $N\to\infty$ and $\tau\to1$, the Hermitian limit, taken such that the global density collapses to the real line, the interval $[-1,1]$ in our case, 
whereas local correlation functions still extend into the complex plane. 
In the elliptic Ginibre ensemble the phenomenon of weak non-Hermiticity  happens at different scales in $N$ in the bulk \cite{FKSlong} and at the soft edge \cite{Bender} of the spectrum. In contrast, in our Coulomb gas living on a finite ellipse this happens on the {\it same} scale in $N$, that is $\tau=1-O(1/N^2)$.  
In our ensemble, with edge we mean the vicinity of the endpoints $\pm1$, and with
bulk we mean the vicinity of interior points of the open interval $(-1,1)$, away from the edges. In view of the fact 
that for $a=O(1)$ the limiting global density of the known Jacobi ensemble 
\cite{FK,NW} diverges like a square root at the endpoints $\pm1$, we expect hard edge behaviour at our edge points. 
For the chiral ensemble \cite{Osborn} the scaling of weak non-Hermiticity in $N$ also agrees with the bulk scaling 
\cite{FKSlong}, which is consistent with our findings. Notice that for any $\tau<1$ 
the foci of our ellipse \eqref{ellipse} are located at $\pm1$ in the interior of the ellipse.

Let us emphasise that our bulk limit is special though, as in this limit the edges of the ellipse become 
close to the real interval $(-1,1)$. Thus our bulk points become squeezed between these edges,  representing 
hard walls, in the vicinity of the interval.  For that reason we may expect that our bulk limit differs from the bulk limit of the Ginibre ensemble. Only when the bulk becomes broader again we recover the Ginibre result, see Subsection \ref{weak-bulk}.

The weak non-Hermiticity limit both in the bulk and at the edge of the spectrum is defined by taking the limit $\tau\to1$ such that 
\begin{equation}
\label{weaklimit}
\frac{1}{\tau} = 1 + \frac{s^2}{2 N^2}, \ \ \ 0 < s < \infty,
\end{equation}
with $N\to\infty$, and the weak non-Hermiticity parameter $s$ is kept fixed\footnote{Note that in \cite{FKSlong} this parameter is typically found to be proportional to $\sim(1-\tau)N$.}. 
For later use we collect the following expressions
\begin{equation}
\tau = \frac{1}{\displaystyle 1+\frac{s^2}{2N^2}}, \ \ \ 
\frac{\tau}{1-\tau}  = \frac{2N^2}{s^2}, \ \ \ 
\frac{\tau}{1+\tau} = \frac{2N^2}{4N^2+s^2}.
\label{tauid}
\end{equation}

Given that the Gegenbauer polynomials can be expressed in terms of the Jacobi polynomials, e.g in \eqref{CP}, it turns out 
that in both the bulk and edge limits the following asymptotic form of 
the general Jacobi polynomials $P^{(\alpha,\gamma)}(z)$ will be useful,  \cite[18.11.5]{NIST}:
\begin{equation}
\label{Jacobiasymptotic}
P^{(\alpha,\gamma)}_n\left(1 - \frac{Z}{2 n^2} \right) \sim n^{\alpha}
\left( \frac{\sqrt{Z}}{2} \right)^{ -\alpha} J_{\alpha}\left(\sqrt{Z}\right), \ \ \ n \rightarrow \infty,
\end{equation}
with fixed real $\alpha$ and $\gamma$, and $Z = X + i Y$ ($X$ and $Y$ are real) kept fixed. Recall that 
the polynomials $P^{(\alpha,\gamma)}(x)$ are orthogonal with respect to the weight $(1-x)^\alpha(1+x)^\gamma$ on $[-1,1]$,  and satisfy the following reflection symmetry:
\begin{equation}
P^{(\alpha,\gamma)}_n(-z)=(-1)^n P^{(\gamma,\alpha)}_n(z),
\label{Psym}
\end{equation}
and that the asymptotic form \eqref{Jacobiasymptotic} zooming into the vicinity of $+1$ is independent of $\gamma$. 

\subsection{Weak non-Hermiticity in the bulk}\label{weak-bulk}

In this subsection we consider the bulk scaling limit in the vicinity of the origin, by rescaling the complex variables inside the kernel (\ref{kernel}) as 
\begin{equation}
\label{bulkvariable}
z_j=x_j + iy_j =  \frac{\hat{z}_j}{N}, \ \ \  j=1,2,
\end{equation}
where $\hat{z}_j=\hat{x}_j+i\hat{y}_j$  (${\hat x}_j$ and $\hat{y}_j$ are real) are kept fixed when $N\to\infty$. We expect that the limiting kernel, after some suitable modification, does not depend on the location in the bulk, and we will check this conjecture with a consistency check in the next Subsection \ref{weak-edge}.

As a short calculation for the scaling limit (given by (\ref{weaklimit}) and (\ref{bulkvariable})) of the pre-factors of 
the kernel in the first line of (\ref{kernel}), that originate from the weight function, we obtain
\begin{equation}
\label{weightlimit}
\lim_{N\to\infty}\left(1 - \frac{2 \tau}{1 + \tau} x_j^2 - \frac{2 \tau}{1 - \tau} y_j^2 \right)^{a/2}
=\left(1 - 4\frac{\hat{y}_j^2}{s^2}\right)^{a/2},
\end{equation}
for $j=1,2$. Here, only the imaginary part of the scaling variable $\hat{z}_j=\hat{x}_j+i\hat{y}_j$ appears.  
From this limit we can read off the domain of the scaling variables ${\hat z}_j$ ($j=1,2$) in the bulk limit:
\begin{equation}
\label{Dbulk}
D_{\rm Bulk} = \left\{ \hat{z}\left|  \frac{s^2}{4}\geq \hat{y}^2\ \ \mbox{and} \ 
-\infty < \hat{x} < \infty
\right. \right\},
\end{equation}
with $\hat{z}=\hat{x}+i\hat{y}$ (${\hat x}$ and ${\hat y}$ are real).
\par
In the kernel (\ref{kernel}) the sum will turn into an integral. Because we split the sum into 
its even and odd parts, let us present the details of this step. For $f_n$ 
some continuous and integrable function depending on $n$ we have
\begin{eqnarray} 
\sum_{n=0}^{N-1}(n+a+1)f_n &=&\sum_{\ell=0}^{\left\lfloor \frac{N-1}{2}\right\rfloor}(2\ell+a+1)f_{2\ell}+
\sum_{\ell=0}^{\left\lfloor \frac{N-2}{2}\right\rfloor}(2\ell+a+2)f_{2\ell+1}
\nonumber\\
&\sim&\frac{N^2}{2} \int_0^1dc \ c\left( f\left(\frac{2\ell}{N}=c\right) +f\left(\frac{2\ell+1}{N}=c\right)\right),
\label{integralsplit}
\end{eqnarray}
in the limit $N \rightarrow \infty$, where $\displaystyle \ell = \left\lfloor n/2 \right\rfloor$. 
We also introduced the integration variable
\begin{equation}
c= \frac{n}{N} = \frac{2\ell}{N} \ {\rm  or} \  \frac{2 \ell + 1}{N} \in [0,1]\ ,
\end{equation}
and use that 
\begin{equation}
\frac{2}{N} \sum_{\ell=0}^{\cal L} \to \int_0^1 dc, \ \ \ \mbox{for}\ \ {\cal L}  = \left\lfloor \frac{N-1}{2} 
\right\rfloor \ {\rm or} \ \left\lfloor \frac{N-2}{2}\right\rfloor.
\end{equation}
\par
For the asymptotic form of the Gegenbauer polynomials 
inside the sum of (\ref{kernel}), we can apply the asymptotic form of the 
Jacobi polynomials (\ref{Jacobiasymptotic}).  As we zoom into the origin with small 
argument of the Gegenbauer polynomials \eqref{bulkvariable},  while
the asymptotic \eqref{Jacobiasymptotic} is in the vicinity of the endpoint, 
we cannot use the standard mapping 
\eqref{CP} of the Gegenbauer polynomials to the symmetric Jacobi polynomials. Fortunately a different map exists, and 
we begin with the even Gegenbauer polynomials. 
Using \cite[18.7.15]{NIST},  we have
\begin{eqnarray}
\label{map1}
C_{2 \ell}^{(a+1)}(x)&=&\frac{(a+1)_\ell}{
\left(1/2\right)_\ell} P_\ell^{(a+\frac12,\ -\frac12)}(2 x^2 - 1)
\nonumber\\
&=&
\frac{\Gamma(\ell+a+1) \ \Gamma(1/2)}{\displaystyle 
\Gamma(a+1) \ \Gamma\left(\ell+\frac12\right)}
(-1)^\ell P_\ell^{(-\frac12,\ a+\frac12)}(1 - 2 x^2 ),
\end{eqnarray}
where $(b)_n=\Gamma(b+n)/\Gamma(b)$ is the Pochhammer symbol.
From (\ref{Jacobiasymptotic}) we thus obtain
\begin{eqnarray}
\lim_{N\to\infty}
\frac{1}{N^a} (-1)^\ell C_{2\ell}^{(a+1)}\left( \frac{\hat{z}}{N}\right)
&=&\frac{\sqrt{\pi} c^a}{2^a \Gamma(a+1)}\lim_{N\to\infty} \ell^{\frac12}
P_\ell^{(-\frac12,\ a+\frac12)}\left(1-2\frac{\hat{z}^2}{N^2}\right)
\nonumber\\
&=& \frac{\sqrt{\pi} c^a }{2^a \Gamma(a+1)} \left( \frac{c \hat{z}}{2} \right)^{\frac12} 
J_{-\frac12}(c \hat{z})
\nonumber\\
&=&\frac{c^a}{2^a \Gamma(a+1)}\cos(c \hat{z}).
\label{C2lasymp}
\end{eqnarray}
Here, $c = 2 \ell/N$ is fixed in the limit $N \rightarrow \infty$, and in 
the last step we have used \cite[8.464.2]{GR}
\begin{equation}
\label{J-1/2}
J_{-\frac12}(z)= \sqrt{\frac{2}{\pi z}}\cos(z).
\end{equation}
The very same steps can be taken for 
the asymptotic form of the odd Gegenbauer polynomials. Using 
\cite[18.7.16]{NIST}, we start from the map
\begin{eqnarray}
\label{map2}
C_{2\ell+1}^{(a+1)}(x)&=&\frac{(a+1)_{\ell+1}}{\left(1/2\right)_{\ell+1}} x P_\ell^{(a+\frac12,\ \frac12)}(2x^2-1)
\nonumber\\
&=&
\frac{\Gamma(\ell+a+2) \ \Gamma(1/2)}{\displaystyle 
\Gamma(a+1) \ \Gamma\left(\ell+\frac32\right)}
(-1)^\ell x P_\ell^{(\frac12,\ a+\frac12)}(1-2x^2).
\end{eqnarray}
Once again (\ref{Jacobiasymptotic}) leads to
\begin{eqnarray}
\lim_{N\to\infty}
\frac{1}{N^a} (-1)^\ell C_{2\ell+1}^{(a+1)}\left( \frac{\hat{z}}{N}\right)
&=&\frac{\sqrt{\pi} c^a}{2^a \Gamma(a+1)} \lim_{N\to\infty}  \frac{{\hat z}}{N} \ell^{\frac12}
P_\ell^{(\frac12,\ a+\frac12)}\left(1-2\frac{\hat{z}^2}{N^2}\right)
\nonumber\\
&=& \frac{\sqrt{\pi} c^a}{2^a \Gamma(a+1)} \left( \frac{c \hat{z}}{2} \right)^{\frac12} 
J_{\frac12}(c \hat{z})
\nonumber\\
&=&\frac{c^a}{2^a \Gamma(a+1)}\sin(c \hat{z}).
\label{C2l+1asymp}
\end{eqnarray}
Here, $c = (2 \ell + 1)/N$ is fixed in the limit $N \rightarrow \infty$, 
and in the last step we have used \cite[8.464.1]{GR}
\begin{equation}
\label{J1/2}
J_{\frac12}(z)= \sqrt{\frac{2}{\pi z}}\sin(z).
\end{equation}

For the Gegenbauer polynomials from the normalisation in the denominator inside the sum of (\ref{kernel}),  
the argument is $1/\tau$. Using \eqref{weaklimit}, we see that we can directly use \eqref{Jacobiasymptotic} together with the standard map \eqref{CP}, valid for both even and odd polynomials alike. 
By analytic continuation of the asymptotic (\ref{Jacobiasymptotic}) to imaginary argument, $Z\to iZ$, we obtain for 
the normalising Gegenbauer polynomial of the scaling variable (\ref{weaklimit})
\begin{equation}
\label{normlimit}
\lim_{n\to\infty} \frac{1}{N^{2a+1}}C^{(a + 1)}_n\left( 1 + \frac{s^2}{2 N^2} \right) =
\frac{\Gamma\left(a +  \frac{3}{2} \right)}{\Gamma(2 a + 2)} \left( 
\frac{2}{cs} \right)^{ a +\frac{1}{2}} I_{a + \frac{1}{2}}(cs),
\end{equation}
with $c=n/N$ fixed.
Here, $I_\alpha(z)$ is the modified Bessel function of the first kind. 

Putting all the above together we obtain the following result 
for the bulk scaling limit of the kernel (\ref{kernel}) around the origin:
\begin{eqnarray}
\label{bulkkernel}
K_{\rm Bulk}(\hat{z}_1,\hat{z}_2)&=&\lim_{N\to\infty}\frac{1}{N^2} K_N\left(\frac{\hat{z}_1}{N},\frac{\hat{z}_2}{N}\right) 
\nonumber\\
&=&\left(1-\frac{4\hat{y}_1^2}{s^2}\right)^{\frac{a}{2}}\left(1-\frac{4\hat{y}_2^2}{s^2}\right)^{\frac{a}{2}}
\frac{1}{\pi s}
\frac{s^{a+\frac12}\Gamma(2a+2)}{\displaystyle 
2^{3a+\frac12}\Gamma\left(a+\frac32\right)\Gamma(a+1)^2}
\nonumber\\
&&\times \int_0^1dc \ \frac{c^{a+\frac12}\left(
\cos(c\hat{z}_1)\cos(c\bar{\hat{z}}_2)+ \sin(c\hat{z}_1)\sin(c\bar{\hat{z}}_2)
\right)}{I_{a + \frac{1}{2}}(cs)} \nonumber\\
&=& \frac{2}{s\pi^{\frac32} \Gamma(a+1)}\left(1-\frac{4\hat{y}_1^2}{s^2}\right)^{\frac{a}{2}}
\left(1-\frac{4\hat{y}_2^2}{s^2}\right)^{\frac{a}{2}}
\int_0^1dc\frac{ \left(cs/2\right)^{a+\frac12}}{I_{a + \frac{1}{2}}(cs)}\cos(c(\hat{z}_1-\bar{\hat{z}}_2)). 
\nonumber \\
\end{eqnarray}
In the second step we have used an addition theorem for the trigonometric functions and (\ref{double}). 
\par
\begin{figure}[H]
\begin{center}
\includegraphics[height=4cm]{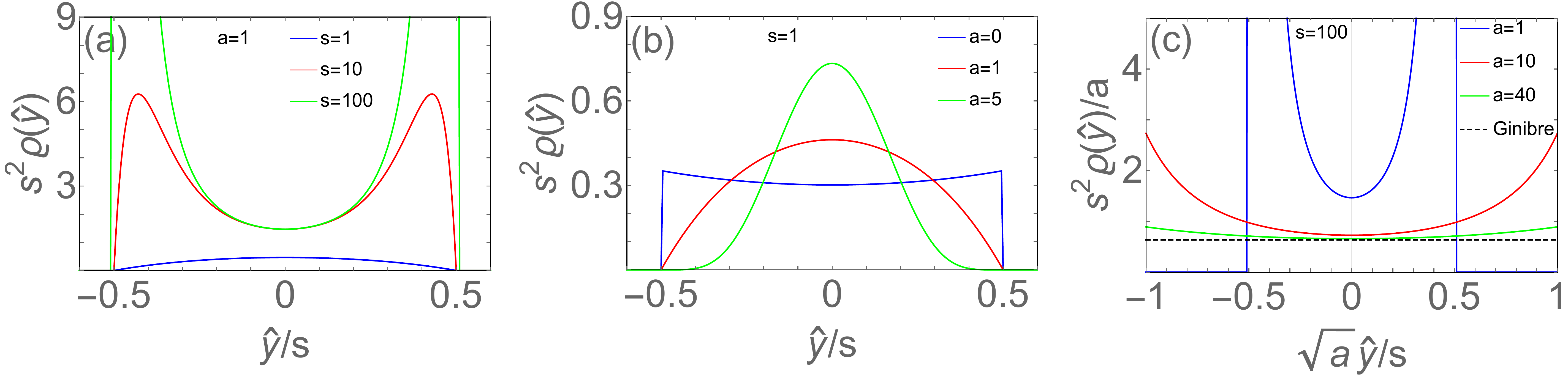}
\caption{\small Microscopic level density~\eqref{dens.bulk} as a function of the 
imaginary part $\hat{y}$ for various charges $a$ of the ellipses 
boundary and for various values of the non-Hermiticity parameter $s$. In 
the plot (a) with $a =1$ and plot (b) with $s=1$ fixed we employ the scaling 
of the strong non-Hermiticity limit with the domain (\ref{Dbulk-strong}), hence, 
the fixed support of $\hat{y}$ is the interval 
$[-1/2,1/2]$. In contrast, we want to illustrate the limit to the 
Ginibre result (dashed straight line on the height $2/\pi$) in the plot 
(c). Therefore, here the size of the support grows with $\sqrt{a}$.}
\label{fig.1}
\end{center}
\end{figure}

The corresponding microscopic level density only depends on the imaginary part, see (\ref{bulkkernel}), 
and reads
\begin{equation}
\label{dens.bulk}
\varrho(\hat{y})=K_{\rm Bulk}({\hat x} + i\hat{y},{\hat x} + i\hat{y}).
\end{equation}
In Fig.~\ref{fig.1} we illustrate the effects of 
the parameters $a$ and $s$. While an increasing non-Hermiticity $s$ 
presses the spectrum away from the real axis to the boundary, see 
Fig.~\ref{fig.1}.a), a growing $a$ results in the opposite effect, cf., 
Fig.~\ref{fig.1}.b). The parameter $a$ represents the charge of the hard 
wall of the boundary of the ellipse leading to a repulsion of the 
particles from the boundary. When both parameter grow large and one 
zooms into the real axis we find the translation invariant bulk 
statistics of the Ginibre ensemble, see Fig.~\ref{fig.1}.c).

The limiting kernel (\ref{bulkkernel}) is a deformation of the sine-kernel in the complex plane. It 
holds inside the domain (\ref{Dbulk}), where the two pre-factors originating from the weight have 
non-negative arguments. We conjecture that the same limiting kernel is found, when we zoom into any 
point $x_0\in(-1,1)$,  and make a bulk scaling limit there,  with an appropriate shift of the weight and rescalings. 
This conjecture is supported by the fact that a similar asymptotic form 
(\ref{bulklimit}) holds in the vicinity of the edge, as the bulk limit of the edge kernel.

We note here that - in addition to the pre-factors stemming from the weight function - the deformed sine-kernel 
in the bulk scaling limit at the origin  (\ref{bulkkernel}) also differs inside the integral 
from what is obtained as a deformed sine-kernel in the weak non-Hermiticity limit of the elliptic 
Ginibre ensemble \cite{FKSlong}, cf. \cite[Eq. (2.22)]{APh} for a comparison in that form. There,  
the pre-factor multiplying cosine is replaced by a simple exponential. 
This difference remains valid  for any fixed value of $a>-1$ as well as  for large arguments, as we will see below.

In the following we will take two limits of the bulk kernel (\ref{bulkkernel}) in order to compare to other known results. We begin with the Hermitian limit as a consistency check.
\par
\medskip
\noindent
(1) \underline{The Hermitian limit $s \rightarrow 0$:}
\par
\medskip
\noindent
In this limit the local bulk kernel is mapped back to the real axis. This can be seen from  the support (\ref{Dbulk}) 
of length $s$ in $\hat{y}$-direction shrinking to zero, leading to $\hat{y}_1, \hat{y}_2 \to0$ in \eqref{bulkkernel}.
For the denominator of the integrand we have the small argument  asymptotic relation of the modified Bessel-function, see e.g. in \cite[8.445]{GR}
\begin{equation}
\label{ssmall}
I_{a + \frac{1}{2}}(c s) \sim \frac{(cs/2)^{a +
  \frac{1}{2}}}{
  \Gamma\left( a + \frac{3}{2} \right)}, \ \ \ s \rightarrow 0.
\end{equation}
Before taking the limit $s\to0$ we have to recall that the ellipse $E$ and $[-1,1]$ are normalised differently, see \eqref{Dnorm} and \eqref{Bdef}. Because from \eqref{tauid} we can read off the constant 
$\displaystyle A\sim \frac{1}{N}\frac{s\pi}{2(a+1)}$, we propose to take the following normalised Hermitian limit
\begin{eqnarray}
\lim_{s\to0} \left. \frac{s\pi}{2(a+1)B} \ K_{\rm Bulk}(\hat{z}_1,\hat{z}_2) \right|_{\hat{y}_1 = \hat{y}_2 = 0} &=&\frac{\pi}{2(a+1)B}
\frac{2 \Gamma\left(a+\frac32\right)}{\pi^{\frac32} \Gamma(a+1)}
\int_0^1dc\cos(c(\hat{x}_1-{\hat{x}}_2))\nonumber\\
&=&\frac1\pi \frac{\sin(\hat{x}_1-{\hat{x}}_2)}{\hat{x}_1-{\hat{x}}_2}
\ .
\label{Ksinlim}
\end{eqnarray}
It results into the well-known universal sine-kernel. It is known to hold for 
the Jacobi ensemble in the bulk of the spectrum \cite{FK}, as well as for other ensembles. 
\par
\medskip
\noindent
(2) \underline{The strong non-Hermiticity limit $s \rightarrow \infty$:} 
\par
\medskip
\noindent
This limit is 
expected to reproduce the limiting kernel at strong non-Hermiticity, when rescaling 
$\tilde{z}_j=\tilde{x}_j+i\tilde{y}_j=\hat{z}_j/s$ for $j=1,2$ (${\tilde x}_j$ and ${\tilde y}_j$ are real). The same mechanism was applied in the elliptic Ginibre ensemble in \cite{FKSlong}.
The corresponding domain \eqref{Dbulk} gets mapped to
\begin{equation}
\label{Dbulk-strong}
D_{\rm Bulk,\ strong} = \left\{ \tilde{z} \left|  \frac{1}{4}\geq \tilde{y}^2\ \ \mbox{and} \ 
- \infty < \tilde{x} < \infty
\right. \right\},
\end{equation}
with $\tilde{z}=\tilde{x}+i\tilde{y}$ (${\tilde x}$ and ${\tilde y}$ are real). It is an infinite strip of 
unit width parallel to the $\tilde{x}$-axis. We obtain the following expression for the limit of the integral in (\ref{bulkkernel}):
\begin{eqnarray}
\label{ja}
\mathcal{J}_a &=& \lim_{s\to\infty}s\int_0^1 dc\frac{\left(cs/2\right)^{a+\frac12}}{I_{a + 
\frac{1}{2}}(cs)}\cos(c(\hat{z}_1-\bar{\hat{z}}_2))
\nonumber\\
&=& \lim_{s\to\infty} \int_0^s dt \frac{\left(t/2\right)^{a+\frac12}}{I_{a + 
\frac{1}{2}}(t)}\cos(t(\tilde{z}_1-\bar{\tilde{z}}_2))
\nonumber\\
&=& \int_0^\infty dt \frac{\left(t/2\right)^{a+\frac12}}{I_{a + 
\frac{1}{2}}(t)}\cos(t(\tilde{z}_1-\bar{\tilde{z}}_2)).
\end{eqnarray}
Here we have changed the integration variable to $t = cs$ . The final answer for the limiting kernel at strong non-Hermiticity on the domain \eqref{Dbulk-strong} thus reads
\begin{eqnarray}
K_{\rm Bulk,strong}(\tilde{z}_1,\tilde{z}_2)&=&
\lim_{s\to\infty} s^2 K_{\rm Bulk}(s\tilde{z}_1,s\tilde{z}_2)\nonumber\\
&=& 
\frac{2}{\pi^{\frac32} \Gamma(a+1)}\left(1- 4\tilde{y}_1^2 \right)^{\frac{a}{2}}
\left(1- 4\tilde{y}_2^2 \right)^{\frac{a}{2}} \int_0^\infty dt \frac{\left(t/2\right)^{a+\frac12}}{I_{a + 
\frac{1}{2}}(t)}\cos(t(\tilde{z}_1-\bar{\tilde{z}}_2)). \nonumber \\ 
\label{strongkernel}
\end{eqnarray}
Although we have derived the kernel \eqref{strongkernel} indirectly via the weak non-Hermiticity limit at the origin, we conjecture it to be universal, after an appropriate shift of the weight away from the origin plus rescalings. 
Because the appropriate Mehler or Poisson formula for the kernel \eqref{kernel} is lacking, when extending 
the sum to infinity\footnote{Notice that a different Poisson kernel exists for the general Jacobi polynomials, cf. \cite{Karp}.},  we have been unable to directly take the strong non-Hermiticity limit.

Notice that the kernel \eqref{strongkernel} does not agree with the Ginibre kernel in the bulk of the spectrum of 
the elliptic Ginibre ensemble. In oder to recover the Ginibre kernel, we need to take the limit $a\to\infty$ 
with a suitable scaling, as explained below. Furthermore, yet another limiting kernel exists, which is obtained when 
imposing a hard edge (at the otherwise soft edge) for the Ginibre ensemble, cf. \cite[Theorem 2.3]{Yacin}. Apparently 
the role of a hard edge differs when imposed for a confining potential as for Ginibre, or for a non-confining potential 
as here, generalising  the Jacobi ensemble.

Let us explain how to recover the Ginibre kernel in the limit $a \rightarrow \infty$. A series expansion \cite[10.25.2]{NIST}
\begin{equation}
\label{series}
I_{a + \frac12}(t) = \left( \frac{t}{2} \right)^{a + \frac12} \sum_{\ell = 0}^\infty \frac{(t^2/4)^\ell}{\ell! 
\ \Gamma\left(\ell + a + \frac32\right)}
\end{equation}
is known for the modified Bessel function. Introducing a new variable ${\hat t} = t/\sqrt{a}$ and 
using the asymptotic relation
\begin{equation}
\frac{\Gamma\left(a + \frac32 \right)}{\Gamma\left(\ell + a + \frac32 \right)} \sim a^{-\ell}, \ \ \ a \rightarrow \infty,
\end{equation}
for a fixed non-negative integer $\ell$, we obtain
\begin{equation}
I_{a + \frac12}(\sqrt{a} {\hat t}) \sim \left( \frac{\sqrt{a} {\hat t}}{2} \right)^{a + \frac12} 
\frac{e^{{\hat t}^2/4}}{\Gamma\left(a + \frac32\right)}, \ \ \ a \rightarrow \infty,
\end{equation}
from (\ref{series}). Here ${\hat t}$ is fixed. We put this asymptotic form into (\ref{ja}) and find
\begin{eqnarray}
\mathcal{J}_a &=& \sqrt{a} \int_0^\infty d{\hat t} \frac{\left(\sqrt{a} {\hat t}/2\right)^{a+\frac12}}{I_{a + 
\frac{1}{2}}(\sqrt{a} {\hat t})}\cos(\sqrt{a} {\hat t}(\tilde{z}_1-\bar{\tilde{z}}_2)) 
\nonumber \\ 
& \sim & \sqrt{a} \ \Gamma\left( a + \frac32 \right) \int_0^\infty d{\hat t} \ e^{- {\hat t}^2/4} 
\cos\left({\hat t} (u_1 - {\bar u}_2) \right) \nonumber \\ 
& = & \sqrt{\pi a} \ \Gamma\left( a + \frac32 \right) e^{- (u_1 - {\bar u}_2)^2},
\end{eqnarray}
where $u_j = \sqrt{a} \tilde {z}_j$ ($j = 1,2$). Then it follows that
\begin{eqnarray}
{\widetilde K}_{Ginibre}(u_1,u_2)&=&
\lim_{a\to\infty} K_{\rm Bulk, \ strong}\left( u_1/\sqrt{a}, u_2/\sqrt{a} \right)/a \nonumber\\
&=& \frac{2}{\pi} \exp\left[ -  |u_1|^2 - |u_2|^2 + 2 u_1 {\bar u}_2 - i
\Im(u_1^2 - u_2^2) \right].
\end{eqnarray}
This kernel is equivalent\footnote{Two kernels are 
equivalent if they agree up to multiplication by $f(u_1)/f(u_2)$ as they yield the same correlation functions 
in \eqref{det}, with $f(u_1) = e^{- i \Im u_1^2}$ here.} to the Ginibre kernel 
$K_{Ginibre}(u_1,u_2)$, presented below.

Though the Ginibre kernel was originally found in \cite{Ginibre} for the Gaussian random matrix model 
(with the kernel (\ref{kernelGin}) in the limit $\tau \rightarrow 0$), it can also be derived from truncated 
unitary random matrices \cite{ZS}. Starting from 
the kernel function (\ref{rs}) of truncated unitary random matrices, one can 
take the aymptotic limit $N \rightarrow \infty$ and obtains
\begin{eqnarray}
\label{rslimit}
K^{truncated}(z_1,z_2)&=& \lim_{N \rightarrow \infty} 
K_N^{truncated}(z_1,z_2) \nonumber \\
&=&
(1 - |z_1|^2)^{\frac{a}{2}} (1 - |z_2|^2)^{\frac{a}{2}} \sum_{n=0}^{\infty} 
\frac{\Gamma(n + a + 2)}{\pi \Gamma(a + 1) \Gamma(n + 1)} 
(z_1 {\bar z}_2)^n \nonumber \\ 
& = & \frac{a + 1}{\pi} \frac{\displaystyle (1 - |z_1|^2)^{\frac{a}{2}} (1 
- |z_2|^2)^{\frac{a}{2}}}{\displaystyle (1 - z_1 {\bar z}_2)^{a+2}},
\end{eqnarray}
for fixed $z_1$ and $z_2$ satisfying $|z_1| < 1$ and $|z_2| < 1$. Introducing 
variables $\displaystyle u_j = \sqrt{\frac{a}{2}} z_j$ ($j = 1,2$) and taking the limit $a \rightarrow 
\infty$, one arrives at the Ginibre kernel
\begin{eqnarray}
K_{Ginibre}(u_1,u_2)&=&
\lim_{a\to\infty} K^{truncated}\left(u_1/\sqrt{a/2},\ u_2/\sqrt{a/2}\right)/(a/2) \nonumber\\
&=& \frac{2}{\pi} \exp\left( - |u_1|^2 - |u_2|^2 + 2 u_1 {\bar u}_2 \right).
\end{eqnarray}

\subsection{Weak non-Hermiticity at the edge}\label{weak-edge}

In this subsection we consider the weak non-Hermiticity limit at the edge of the spectrum. Because the Gegenbauer polynomials have parity, without loss of generality we magnify the region around the focus at $+1$, in choosing the scaling 
\begin{equation}
\label{Zedge}
z_j =  1 - \frac{Z_j}{2 N^2}, \ \ \ j=1,2,
\end{equation}
together with the weak non-Hermiticity  limit \eqref{weaklimit}. Here, the complex numbers $Z_j = X_j + i Y_j$ are fixed ($X_j$ and $Y_j$ are real). In this limit the pre-factors of the kernel \eqref{kernel} from the weight turn into
\begin{equation}
\label{limit1}
\left(1 - \frac{2 \tau}{1 + \tau} x_j^2 - \frac{2 \tau}{1 - \tau} 
y_j^2 \right)^{a/2} \sim N^{-a} \left(\frac{s^2}{4} + X_j - \left( \frac{Y_j}{s}
\right)^2 \right)^{a/2},
\end{equation}
in the limit $N \rightarrow \infty$ as \eqref{weaklimit} and \eqref{Zedge}.
Once again we keep the parameter $a$ fixed in this limit. 
Eq. (\ref{limit1}) implies that the limiting domain of the
scaled particle positions $(X_j,Y_j)$ becomes the  parabolic domain
\begin{equation}
\label{para}
D_{\rm Edge} = \left\{(X,Y) \left| X \geq \left( \frac{Y}{s}
\right)^2 - \frac{s^2}{4} \right. \right\},
\end{equation}
which is a magnified part around the right focus of the ellipse, that is the right endpoint of $[-1,1]$.

The pre-factor of the sum in the second line of (\ref{kernel}) is easily evaluated by using (\ref{tauid}), to give
\begin{equation}
\label{prefactor}
\frac{2 \tau}{\pi \sqrt{1 - \tau^2}} =\frac{2}{\pi} \sqrt{\frac{\tau}{1-\tau}\frac{\tau}{1+\tau}}\sim\frac{2N}{s\pi}.
\end{equation}
Due to the relation \eqref{CP} of the Gegenbauer polynomials to the symmetric Jacobi polynomials, and their asymptotic form \eqref{Jacobiasymptotic} in the vicinity of unity, 
we find the following asymptotic relation,
\begin{eqnarray}
\label{limit2}
C^{(a + 1)}_n(z_j) 
=C^{(a + 1)}_n\left( 1 - \frac{Z_j}{2 N^2} \right) \sim N^{2 a + 1} \frac{\displaystyle 
\Gamma\left(a +  \frac{3}{2} \right)}{\Gamma(2 a + 2)} \left( 
\frac{\sqrt{Z_j}}{2 c} \right)^{- a - \frac{1}{2}} 
J_{a + \frac{1}{2}}\left(c\sqrt{Z_j}\right).\ \ 
\end{eqnarray}
Because the limit of the squared norms does not depend on the point we magnify, 
we may use again the asymptotic \eqref{normlimit} from the previous subsection.

Inserting (\ref{limit1}), \eqref{prefactor}, (\ref{limit2}) and (\ref{normlimit}) together in (\ref{kernel}), 
and replacing the sum by an integral, yields the following asymptotic formula for the limiting kernel at the edge 
\begin{eqnarray}
\label{edgekernel}
K_{\rm Edge}(Z_1,Z_2) &=& 
 \lim_{N \rightarrow \infty} \frac{1}{4N^4} K_N(z_1,z_2) \nonumber \\ 
 & = & 
\frac{1}{4\sqrt{\pi} \Gamma(a + 1)} \left( \frac{s}{2}
\right)^{a - \frac{1}{2}}  
\left(\frac{s^2}{4} + X_1 - \left(\frac{Y_1}{s} \right)^2  
\right)^{\frac{a}{2}} \left(\frac{s^2}{4} + X_2 - \left(\frac{Y_2}{s} \right)^2  
\right)^{\frac{a}{2}} \nonumber \\ 
&& \times  \left( \sqrt{ Z_1 {\bar Z_2}} \right)^{- a - \frac{1}{2}} 
\int_0^1 dc \ \frac{c^{a + \frac{3}{2}}}{I_{a + \frac{1}{2}}(c s)} 
J_{a + \frac{1}{2}}\left(c \sqrt{Z_1}\right) 
J_{a + \frac{1}{2}}\left(c \sqrt{\bar{Z_2}}\right),
\end{eqnarray}
with a fixed $a > -1$. This limiting kernel is a deformation of the Bessel-kernel into the complex plane, 
holding inside the domain \eqref{para} where the two pre-factors from the weight have non-negative arguments. 
From symmetry  the same limiting kernel is obtained at the left edge of the ellipse. Not only the pre-factors 
from the weight but also the pre-factor in the integrand inversely proportional to the modified  $I$-Bessel 
function differs from the pre-factor of the deformed Bessel-kernel of the chiral ensemble \cite{Osborn}, given 
by an exponential. There, $a+\frac12=\nu$, and for integer values it corresponds to the number of zero-modes therein.  This difference remains valid for any fixed values $a>-1$, and shows the influence of the boundary. 
It pertains also for large arguments, as we will see below.  We expect that the limiting edge-kernel \eqref{edgekernel} is also universal.

\begin{figure}[H]
\begin{center}
\includegraphics[height=5cm]{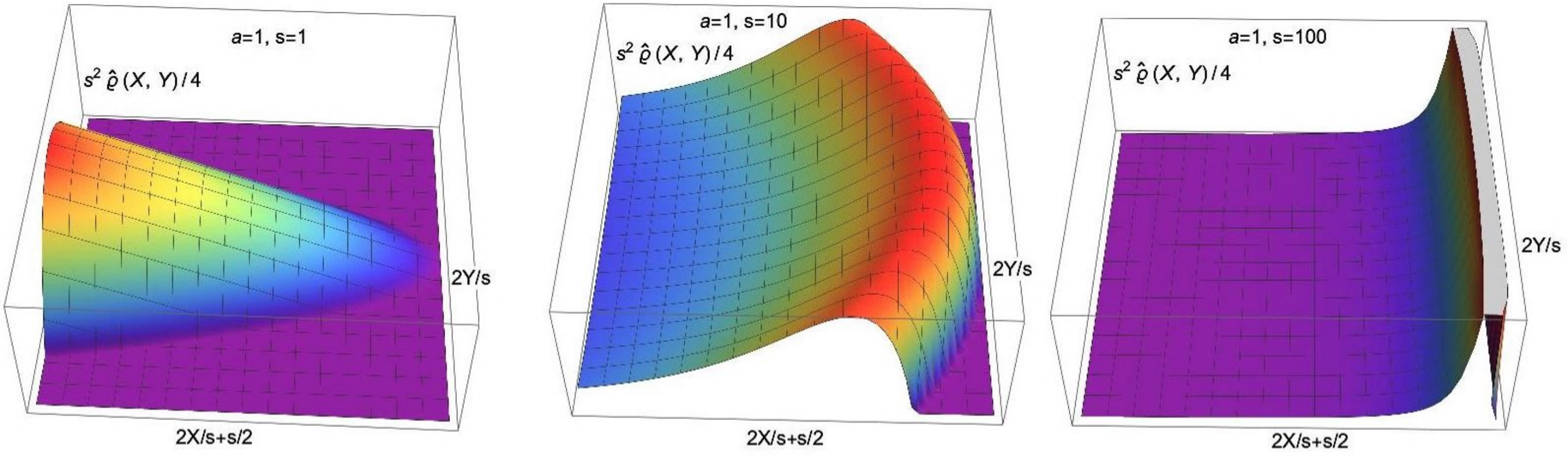}
\caption{\small The rescaled microscopic level density~\eqref{dens.edge} 
$s^2 \hat{\varrho}(X,Y)/4$ at the edge for 
increasing non-Hermiticity $s=1,10,100$ (from left to right) at fixed 
charge $a=1$ of the boundary. The color coding of the graph highlights 
the height of the function. The scaling of the real and imaginary parts 
are those of the strong non-Hermiticity limit, see \eqref{tildescale}.}
\label{fig.2}
\par\bigskip
\end{center}
\end{figure}

\begin{figure}[H]
\begin{center}
\includegraphics[height=5cm]{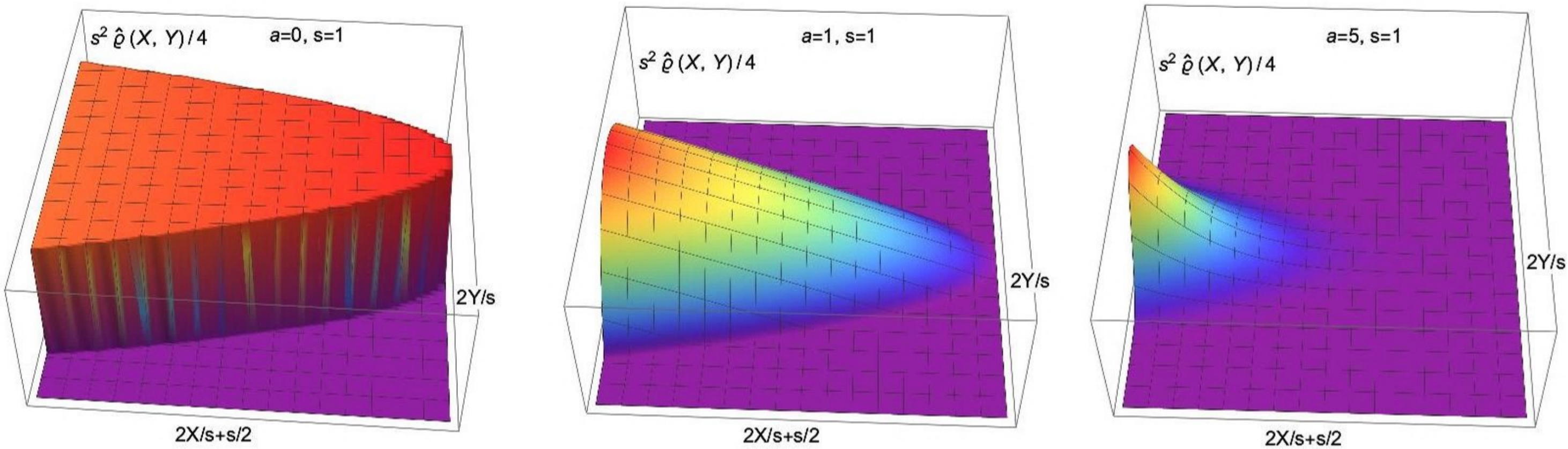}
\caption{\small The rescaled microscopic level density~\eqref{dens.edge} 
$s^2 \hat{\varrho}(X,Y)/4$ at the edge for 
increasing charge $a=0,1,5$ (from left to right) at fixed 
non-Hermiticity $s=1$. Again we have employed the 
scaling~\eqref{tildescale}, see also Fig.~\ref{fig.2}.}
\label{fig.3}
\end{center}
\end{figure}

Again we define a microscopic density which depends this time on both 
the real and imaginary parts, due to the loss of translation invariance, i.e.,
\begin{equation}
\label{dens.edge}
\hat{\varrho}(X,Y)=K_{\rm Edge}(X+iY,X+iY).
\end{equation}
Its dependence on an increasing non-Hermiticity $s$ and an increasing 
charge $a$ is illustrated in Figs.~\ref{fig.2} and~\ref{fig.3}, 
respectively. Note that the positive direction of the horizontal axis is 
the direction to the left to reflect the position of the edge where we 
zoom into the spectrum. At the edge we have a similar picture compared to the 
microscopic bulk regime. The spectrum lies in a constant competition 
between $s$, which tries to spread and squeeze it into the boundary, and 
$a$, which creates a repulsion from exactly the same boundary.

Below we will take two limits of the kernel \eqref{edgekernel} to compare with known asymptotic kernels 
in random matrix theory, the Hermitian and strong non-Hermiticity limit. In addition we take a third 
limit of large argument, that brings us back to the result in the bulk from the previous subsection.
\par
\medskip
\noindent
(1) \underline{The Hermitian limit $s \rightarrow 0$:}
\par
\medskip
\noindent
In this limit, we can see from the domain \eqref{para} that it requires $Y_j = 0$, and the real parts  
are confined to the half line, $X_j \geq 0$. For the normalisation of this Hermitian limit we follow \eqref{Ksinlim}, and 
for the pre-factor inside the integral in \eqref{edgekernel} we may use again the asymptotic \eqref{ssmall}. This leads to the following result:
\begin{eqnarray}
\label{besselkernel}
& & \lim_{s \rightarrow 0}  \left. \frac{s \pi}{2(a+1)B} K_{\rm Edge}(Z_1,Z_2) \right|_{X_{1,2}\geq0, \ Y_{1,2} = 0} \nonumber \\ 
& = & 
\frac14
\left( X_1 X_2\right)^{- \frac{1}{4}} 
\int_0^1 dc \ c \ J_{a + \frac{1}{2}}\left(c \sqrt{X_1} \right) 
J_{a + \frac{1}{2}}\left(c \sqrt{X_2} \right),
\end{eqnarray}
with a fixed $a>-1$. This reproduces a well-known universal result, the Bessel-kernel, derived for the symmetric Jacobi ensemble of random Hermitian matrices \cite{NW} with weight \eqref{Jacobi}. Note that the non-constant 
pre-factor $(X_1 X_2)^{-1/4}$ is cancelled in the expressions of  the correlation functions, when we make variable transformations $X_j \mapsto X_j^2$.
\par
\medskip
\noindent
(2) \underline{The strong non-Hermiticity limit $s \rightarrow \infty$}:
\par
\medskip
\noindent
Let us next consider the opposite limit $s \rightarrow \infty$, to obtain the limiting kernel at strong non-Hermiticity. For that purpose, we introduce new scaling 
variables
\begin{equation}
\label{tildescale}
{\tilde X}_j = \frac{2}{s} X_j + \frac{s}{2}, \ \ \ {\tilde Y}_j = 
\frac{2}{s} Y_j, 
\end{equation}
where we keep ${\tilde X}_j$ and ${\tilde Y}_j$ fixed when taking the limit $s \rightarrow 
\infty$. In terms of these new variables the determining equation for the domain \eqref{para} becomes 
$\displaystyle \frac{s}{2}\tilde{X}_j\geq \frac{\tilde{Y}_j^2}{4}$. 
Thus in the limit the scaled particle 
positions $({\tilde X}_j, {\tilde Y}_j)$ are confined to the half plane, 
that is $0 \leq  {\tilde X}_j < \infty$ and $-\infty < {\tilde Y}_j < \infty$. 
Now we use the asymptotic formula \cite{szego} for $u 
\rightarrow \infty$,
\begin{equation}
\label{jcos}
J_b\left(u z \right) \sim \left( \frac{2}{\pi u z} 
\right)^{1/2} \cos\left(u z - \frac{\pi}{2} b - 
\frac{\pi}{4} \right),
\end{equation}
for a fixed real index $b$ and a fixed complex $z$,  to obtain
\begin{eqnarray}
\left( \sqrt{Z}_j \right)^{-a - \frac{1}{2}} J_{a + \frac{1}{2}}\left( c \sqrt{Z}_j
\right) 
\sim
\left( \frac{s}{2} \right)^{-a - \frac{1}{2}} \left( \pi c s \right)^{-1/2} 
{\rm exp}\left[ \frac{c s}{2} \left(1 - \frac{1}{s} \left( 
{\tilde X}_j + i {\tilde Y}_j \right) \right) \right],
\end{eqnarray}
for $s \rightarrow \infty $. Together with the large-$s$ asymptotic for the modified Bessel functions, 
cf. \cite[8.451.5]{GR}, 
\begin{equation}
I_{a + \frac{1}{2}}\left( c s \right) \sim  
\left( 2 \pi c s \right)^{-1/2} e^{cs},
\ \ \ s \rightarrow \infty\ ,
\end{equation}
valid for any fixed $a$, it then follows for the scaling \eqref{tildescale} that
\begin{eqnarray}
\label{sinfinity}
K_{\rm Edge,strong}(\tilde{Z}_1,\tilde{Z}_2)
&=& \lim_{s \rightarrow \infty} \frac{s^2}{4} K_{\rm Edge}(Z_1,Z_2) \nonumber \\ 
& = & \frac{ \left({\tilde X}_1 {\tilde X}_2\right)^{a/2}}{4 \pi \Gamma(a + 1)}  \int_0^1 dc \ c^{a + 1} \exp\left[ -\frac{c}{2} ({\tilde
  X}_1 + {\tilde X}_2) - i \frac{c}{2}({\tilde Y}_1 - {\tilde Y}_2) \right], \nonumber \\ 
\end{eqnarray}
with a fixed $a>-1$. This limiting kernel is not new and, as we will show below, agrees with the kernel found for truncated unitary matrices \cite{ZS} in what the authors call weakly non-unitary limit. What we call strongly non-Hermitian here is to be  understood in the sense that by taking the limit $s\to\infty$ we reestablish rotational invariance.

Starting directly from the kernel of the truncated unitary matrix ensemble \eqref{rs}, we may introduce scaled real variables ${\hat X}_j$ and ${\hat Y}_j$ that remain fixed when $N\to\infty$,
\begin{equation}
z_j = 1 - \frac{{\hat X}_j}{2 N} - i \frac{{\hat Y}_j}{2 N},
\end{equation} 
magnifying the edge region of the unit circle at unity. Then, we obtain
\begin{eqnarray}
\lim_{N \rightarrow \infty} \frac{1}{4 N^2} K_N^{truncated}(z_1,z_2) 
& = & \frac{\left({\hat X}_1{\hat X}_2\right)^{a/2}}{4 \pi \Gamma(a + 1)}  \int_0^1 
dc \ c^{a + 1} \exp\left[-\frac{c}{2} ({\hat
  X}_1 + {\hat X}_2) - i \frac{c}{2}({\hat Y}_1 - {\hat Y}_2) \right].
\nonumber \\ 
\end{eqnarray}
It is in agreement with 
the asymptotic formula (\ref{sinfinity}), and the scaled density $\rho(\tilde{Z}_1)=K_{\rm Edge,strong}(\tilde{Z}_1,\tilde{Z}_1)$ agrees with the density computed in \cite[Eq. (21)]{ZS}. 
\par
\medskip
\noindent
(3) \underline{The bulk limit:}
\par
\medskip
\noindent
It is known that, in taking the large argument limit, the correlations at the edge get mapped back to the correlations in the bulk, see e.g. \cite{ForresterHonner}. Thus this limit will allow us to check our conjecture that a similar 
asymptotic form to  the kernel \eqref{bulkkernel}  is valid in the entire bulk.

Let us therefore introduce scaled complex variables ${\hat z}_j =
 {\hat x}_j + i {\hat y}_j$ for the  arguments of the edge kernel (\ref{edgekernel}) as
\begin{equation}
\label{kappa}
Z_j = \kappa h - 2 \sqrt{h} {\hat z}_j,
\end{equation}
where $\kappa>0$  and $\hat{z}_j$ remain fixed, and
we will take the limit of 
$h$ positive to become large, $h \rightarrow \infty$.  In these variables the defining equation for the domain \eqref{para} with $Z = X + i Y = \kappa h - 2 \sqrt{h} {\hat z}$ becomes 
\begin{equation}
\kappa h -2 \sqrt{h}\hat{x}\geq \frac{4h\hat{y}^2}{s^2}-\frac{s^2}{4}\ ,
\end{equation}
leading to the domain
\begin{equation}
D_{\rm Bulk} = \left\{ \hat{z} \left|  \frac{s^2}{4} \kappa \geq \hat{y}^2
\right. \ \ \mbox{and} \ 
-\infty < \hat{x} < \infty
\right\},
\end{equation}
where  ${\hat z} = {\hat x} + i {\hat y}$. 

For the scaling \eqref{kappa} we can readily  see that
\begin{equation}
\sqrt{Z_j} \sim \sqrt{\kappa h}  - \frac{{\hat z}_j}{\sqrt{\kappa}}, \ \ \ h \rightarrow 
\infty.
\end{equation}
Then, we can utilize (\ref{jcos}) to find that 
\begin{eqnarray}
J_{a + \frac{1}{2}} \left( c \sqrt{Z_j} \right) \sim 
\left( \frac{2}{c \pi \sqrt{\kappa h}} \right)^{1/2} 
\cos\left( c \sqrt{\kappa h} - \frac{c {\hat z}_j}{\sqrt{\kappa}} - \frac{\pi}{2} a - \frac{\pi}{2} 
\right), \ \ \ h \rightarrow \infty. 
\end{eqnarray}
Putting the above asymptotic results for the scaling \eqref{kappa} together in (\ref{edgekernel}), we obtain
\begin{eqnarray}
\label{bulklimit}
K_{\rm Bulk}({\hat z}_1,{\hat z}_2) &=& \lim_{h \rightarrow \infty} 4 h K_{\rm Edge}(Z_1,Z_2) 
\nonumber \\ 
& = & \frac{2}{s\pi^{\frac32} \Gamma(a+1) \kappa^{a + 1}}\nonumber\\
&&\times
\left(\kappa -\frac{4\hat{y}_1^2}{s^2}\right)^{\frac{a}{2}}\left(\kappa -\frac{4\hat{y}_2^2}{s^2}\right)^{\frac{a}{2}}
\int_0^1dc\frac{\left(cs/2\right)^{a+\frac12}}{I_{a + \frac{1}{2}}(cs)}
\cos\left(\frac{c}{\sqrt{\kappa}}(\hat{z}_1-\bar{\hat{z}}_2)\right), \nonumber \\ 
\end{eqnarray}
which is similar to the asymptotic kernel (\ref{bulkkernel}) computed at the origin, in agreement 
with our conjecture.

\section{Global correlations for unit weight $w(z)=1$
}\label{weight1}
\setcounter{equation}{0}
\renewcommand{\theequation}{5.\arabic{equation}}

In this section we will look at global correlation functions in the interior region (global regime) of the ellipse. 
Note that most of the $N$ particles are concentrated in the vicinity of the edge of the ellipse due to the repulsion among them, and 
that only a negligibly small portion of the particles exist in the interior. In the simplest case of an unweighted ellipse, 
that is with weight  $w(z) = 1$ corresponding to $a = 0$, we are able to derive the  global asymptotic formulas for the 
correlation functions in the limit $N \rightarrow \infty$, which are valid in the whole interior of the ellipse. 

Assuming $E\subset \mathbb{C}$ is a simply connected domain, $t$ is a 
fixed point in $E$, and $F$ is the conformal mapping (the Riemann map) 
of $E$ onto the unit disc $D$, normalised by the conditions $F(t)=0$ and 
$F'(t)>0$. As is well-known, these conditions determine $F$ uniquely.

Then, the following theorem \cite[p.33]{Gaier} establishes the 
relationship between the Bergman kernel (called $K_ {\rm global}$ below) 
and the Riemann 
map
\par
\bigskip
\noindent
\begin{thm} (unweighted case) 
\par
\noindent
The conformal mapping $F$ and the Bergman 
kernel function $K_{global}$ of
$E$ are related as follows:
\begin{equation}
K_{global}\left(z,\bar{t}\right)=\frac{1}{\pi}F'(z)F'(t) \quad \text{and} \quad 
F'(z)=\sqrt{\frac{\pi}{K_{global}\left(t,\bar{t}\right)}}K_{global}\left(z,\bar{t}\right) \quad \text{for}\quad 
z \in E.
\end{equation}
\end{thm}
\par
\bigskip
In particular when $F$ is the Riemann mapping of the ellipse into unit 
disk, this is cumbersome, a first attempt for the Chebyshev polynomials 
of the second kind was made in \cite{Nehari}. However, our representation below 
will be somewhat more explicit, allowing for a consistency check in the 
rotationally symmetric limit, but we do not expect further 
simplification.

When setting $a=0$ the Gegenbauer polynomials reduce to the Chebyshev polynomials of the second kind, $U_n(x ) = C^{(1)}_n(x)$. Prior to taking the large-$N$ limit we  introduce the
rescalings $z_j \mapsto z_j/\sqrt{2 \tau}$ ($j=1,2$), thus mapping the ellipse \eqref{ellipse} to
\begin{equation}
\label{ellipse/t}
E_{\rm rescaled} = \left\{z=x+iy \left| \frac{x^2}{1 + \tau} + \frac{y^2}{1 - \tau}\leq 1 \right. \right\}\ ,\quad 0 < \tau < 1\ .
\end{equation}
This is done in order to be able to take the limit of maximal Hermiticity $\tau\to0$ at the end of the calculation as a consistency check.

Setting $a = 0$ and rescaling the arguments, the kernel function (\ref{kernel}) takes the form
\begin{equation}
K_N\left(\frac{z_1}{\sqrt{2 \tau}},\frac{z_2}{\sqrt{2 \tau}}\right) 
= \frac{2 \tau}{\pi \sqrt{1 - \tau^2}} \sum_{n=0}^{N-1} \frac{n + 1}{\displaystyle 
U_n(1/\tau)} U_n\left(\frac{z_1}{\sqrt{2 \tau}}\right) U_n\left(\frac{{\bar z}_2}{\sqrt{2 \tau}}\right).
\end{equation}
We introduce a complex variable $\omega$ and a real variable $v$ as
\begin{equation}
\label{wv}
\frac{z}{\sqrt{2 \tau}} = \frac{1}{2} \left( \omega + \frac{1}{\omega} \right), \ \ \ 
\frac{1}{\tau} = \frac{1}{2} \left( v^2 + \frac{1}{v^2} \right),
\end{equation}
with
\begin{equation}
\label{interior}
1 \leq |\omega| < v\ .
\end{equation}
This implies that $z$ is in the interior of the ellipse (\ref{ellipse/t}). The parametrisation \eqref{wv}, also 
called Joukowsky map, allows to simplify the Chebyshev polynomials $U_n$, and 
we have \cite{MH}
\begin{equation}
\label{Urel}
U_n\left( \frac{z}{\sqrt{2 \tau}} \right) = \frac{\omega^{n + 1} - \omega^{-n-1}}{\omega - \omega^{-1}}, \ \ \
U_n\left( \frac{1}{\tau} \right) = \frac{v^{2 n + 2} - v^{- 2 n - 2}}{v^2 - v^{-2}}.
\end{equation}
Putting these relations into the kernel, we obtain
\begin{eqnarray}
 K_N\left(\frac{z_1}{\sqrt{2 \tau}},\frac{z_2}{\sqrt{2 \tau}}\right) 
&=& \frac{4}{\pi} \frac{1}{(\omega_1 - \omega_1^{-1}) ({\bar \omega}_2 - {\bar \omega}_2^{-1}) } 
\nonumber \\ 
& & \times \sum_{n=0}^{N-1} (n+1)\frac{(\omega_1^{n + 1} - \omega_1^{- n- 1}) 
({\bar \omega}_2^{n + 1} - {\bar \omega}_2^{- n - 1})}{v^{2 n + 2} - v^{- 2 n - 2}},
\end{eqnarray}
where
\begin{equation}
\frac{z_1}{\sqrt{2 \tau}} = \frac{1}{2} \left( \omega_1 + \frac{1}{\omega_1} \right), \ \ \ 
\frac{z_2}{\sqrt{2 \tau}} = \frac{1}{2} \left( \omega_2 + \frac{1}{\omega_2} \right),
\end{equation}
with $1 \leq |\omega_1| < v$ and $1 \leq |\omega_2| < v$. The sum can be rewritten as
\begin{eqnarray}
&&K_N\left(\frac{z_1}{\sqrt{2 \tau}},\frac{z_2}{\sqrt{2 \tau}}\right) 
= \frac{4}{\pi} \frac{1}{(\omega_1 - \omega_1^{-1}) ({\bar \omega}_2 - {\bar \omega}_2^{-1}) } 
\nonumber \\ 
&& \times 
\sum_{j=0}^{\infty} \left. \frac{\partial}{\partial \lambda} \right|_{\lambda = 1}
 \sum_{n=0}^{N-1} \left( (\xi_j \omega_1 {\bar \omega}_2)^{n+1} - (\xi_j \omega_1/{\bar \omega}_2)^{n+1} 
- (\xi_j {\bar \omega}_2/\omega_1)^{n+1} + (\xi_j/(\omega_1 {\bar \omega}_2))^{n+1} \right) ,
\nonumber \\
\end{eqnarray}
by introducing the auxiliary  variable
\begin{equation}
\xi_j = \frac{\lambda}{v^{2(1 + 2 j)}}.
\end{equation}
The differential operator
\begin{equation}
\left. \frac{\partial}{\partial \lambda} \right|_{\lambda = 1}
\end{equation}
means putting $\lambda = 1$, after taking a derivative with respect to $\lambda$. 
We can now evaluate the sums over $n$ as finite geometric series, and find
\begin{eqnarray}
K_N\left(\frac{z_1}{\sqrt{2 \tau}},\frac{z_2}{\sqrt{2 \tau}}\right) 
&=& \frac{4}{\pi} \frac{1}{(\omega_1 - \omega_1^{-1}) ({\bar \omega}_2 - {\bar \omega}_2^{-1}) } 
\nonumber \\ 
& \times & 
\sum_{j=0}^{\infty} \left. \frac{\partial}{\partial \lambda} \right|_{\lambda = 1}
\left[ (\xi_j \omega_1 {\bar \omega}_2) 
\frac{1 - (\xi_j \omega_1 {\bar \omega}_2)^N}{1 - (\xi_j \omega_1 {\bar \omega}_2)}
 - (\xi_j \omega_1/{\bar \omega}_2) \frac{1 - (\xi_j \omega_1/{\bar \omega}_2)^N}{1 - (\xi_j \omega_1/{\bar \omega}_2)} \right.
\nonumber \\ 
& & \left. - (\xi_j {\bar \omega}_2/\omega_1) \frac{1 - (\xi_j {\bar \omega}_2/\omega_1)^N}{1 - (\xi_j {\bar \omega}_2/\omega_1)}
 + (\xi_j/(\omega_1 {\bar \omega}_2)) \frac{1 - (\xi_j/(\omega_1 {\bar \omega}_2))^N}{1 - (\xi_j/(\omega_1 {\bar \omega}_2))} 
\right]. \nonumber \\
\end{eqnarray}
Because of $1 \leq |\omega_1|<v$ and $1 \leq |\omega_2|<v$ , we observe that for all $j$ 
\begin{equation}
\left| \omega_1^{\pm1}\bar{\omega}_2^{\pm1}/v^{2(1+2j)}\right|<1 \ .
\end{equation}
Thus we can take the limit $N \rightarrow \infty$ (with $\tau$ fixed) to obtain
\begin{eqnarray}
K_{\rm global}(z_1,z_2)
&=& \lim_{N \rightarrow \infty} \frac{1}{2 \tau} K_N\left(\frac{z_1}{\sqrt{2 \tau}},\frac{z_2}{\sqrt{2 \tau}}\right) 
\nonumber \\ & = & \frac{2}{\pi \tau} \frac{1}{(\omega_1 - \omega_1^{-1}) 
({\bar \omega}_2 - {\bar \omega}_2^{-1}) } 
\sum_{j=0}^{\infty} \left. \frac{\partial}{\partial \lambda} \right|_{\lambda = 1}
\left[
\frac{\xi_j \omega_1 {\bar \omega}_2}{1 - (\xi_j \omega_1 {\bar \omega}_2)}
\right.
\nonumber \\ 
&& \quad\quad
\left.
 - \frac{\xi_j \omega_1/{\bar \omega}_2}{1 - (\xi_j \omega_1/{\bar \omega}_2)} 
- \frac{\xi_j {\bar \omega}_2/\omega_1}{1 - (\xi_j {\bar \omega}_2/\omega_1)}
 + \frac{\xi_j/(\omega_1 {\bar \omega}_2)}{1 - (\xi_j/(\omega_1 {\bar \omega}_2))} 
\right].\ \ \ \ 
\end{eqnarray}
Taking the derivative with respect to $\lambda$ yields
\begin{eqnarray}
\label{global}
K_{\rm global}(z_1,z_2)
&=& \frac{2}{\pi \tau} \frac{1}{(\omega_1 - \omega_1^{-1}) ({\bar \omega}_2 - {\bar \omega}_2^{-1}) } 
\nonumber \\ 
&& \times  \sum_{j=0}^{\infty} \left[
\frac{\eta_j \omega_1 {\bar \omega}_2}{(1 - (\eta_j \omega_1 {\bar \omega}_2))^2}
- \frac{\eta_j \omega_1/{\bar \omega}_2}{(1 - (\eta_j \omega_1/{\bar \omega}_2))^2} \right.
\nonumber \\ & & 
\quad\quad\left.
- \frac{\eta_j {\bar \omega}_2/\omega_1}{(1 - (\eta_j {\bar \omega}_2/\omega_1))^2}
+ \frac{\eta_j/(\omega_1 {\bar \omega}_2)}{(1 - (\eta_j/(\omega_1 {\bar \omega}_2)))^2} 
\right],\quad 
\end{eqnarray}
with
\begin{equation}
\eta_j = \frac{1}{v^{2(1 + 2 j)}}.
\end{equation}
This is the limiting kernel on a global scale, valid in the entire interior of the ellipse \eqref{ellipse/t}. 
Because of \eqref{det} that remains valid in this limit, it determines all $k$-point correlation functions on 
a global scale. At present we are only able to derive such a global asymptotic formula for the simplest case $a = 0$.  
In Appendices \ref{BChebyshev} and \ref{CChebyshev} we present a similar analysis for the Chebyshev polynomials of the 
first and third kind, with a non-flat measure on the ellipse.

To get an impression of the $\tau$-dependence of the kernel, we consider 
the origin $z = 0$. Here, we can use the relations 
$\omega {\bar \omega} = 1$ and $\omega /{\bar \omega} = -1$ from \eqref{wv}, to obtain
\begin{equation}
K_{\rm global}(0,0) = \frac{2}{\pi \tau v^2} 
\sum_{j=0}^{\infty}v^{4 j}  \frac{\displaystyle  v^{8 j} + \frac{1}{v^4} }{
\displaystyle \left( v^{8 j} - \frac{1}{v^4} \right)^2},
\end{equation}
with the relation between $v$ and $\tau$ from \eqref{wv}.

It is not justified to take the weak non-Hermiticity limit ($N \rightarrow \infty$ with 
the scaling (\ref{weaklimit})) of (\ref{global}) because of the restriction (\ref{interior}) being violated, which was crucial for our analysis above. 
In the opposite limit of maximal non-Hermiticity $\tau \rightarrow 0$, which due to $1<v$ and \eqref{wv} implies that $v \rightarrow \infty$, we introduce the rescaled variables
\begin{equation}
\label{ozeta}
\omega_j = v \zeta_j, \ \ \ j=1,2,
\end{equation}
with $\zeta_j$ fixed. Then, it follows that $z_j \sim \zeta_j$ in the limit $v \rightarrow \infty$.
Moreover, in the sum of (\ref{global}) only the first term $\eta_j \omega_1 
{\bar \omega}_2/(1 - (\eta_j \omega_1 {\bar \omega}_2))^2$ with $j = 0$ 
survives in the limit $v \rightarrow \infty$. We accordingly obtain for the global limiting kernel in the rotationally symmetric case
\begin{equation}
\label{Krot}
\lim_{\tau \rightarrow 0} K_{\rm global}(z_1,z_2)
= \frac{1}{\pi} \frac{1}{(1 - z_1 {\bar z_2})^2}, 
\end{equation}
when $|z_j| < 1$ ($j = 1,2$) according to \eqref{ellipse/t}. It is in agreement with the known result for the radially 
symmetric weight (\ref{truncated}) with $a = 0$, as can be easily seen from the limit (\ref{rslimit}) of the corresponding 
kernel in (\ref{rs}).

\section{Summary}
\label{sum}
\setcounter{equation}{0}
\renewcommand{\theequation}{6.\arabic{equation}}

In this paper (including Appendices) we have introduced three new families (two were studied in Appendix A) 
of Coulomb gases in two dimensions at the specific temperature $\beta=2$, that are constrained to a hard-walled 
ellipse whose boundary is charged as well, and which repels a finite number of particles inside. In some examples 
in Appendices B and C the potentials include singularities at the foci of the ellipse. These results were made 
possible using the technique of planar orthogonal polynomials on that domain, together with newly derived 
orthogonality results for the classical polynomials of Gegenbauer and Jacobi type from a companion paper \cite{ANPV}. 

We have discussed the local correlation functions in large-$N$ limit, at weak and strong non-Hermiticity 
in the bulk and at the edge of the spectrum, by determining the corresponding limiting kernels. We found several 
new universality classes of deformed sine- and Bessel-kernels in the complex plane, that all showed the influence 
of the edges. Several different families led to the same limiting kernel, thus underlying their conjectured universality. 
In the Hermitian limit we could recover the universal sine- and Bessel-kernel of the Jacobi ensemble. At strong 
non-Hermiticity we were led back to the corresponding limiting kernel of the ensemble of Gaussian and 
truncated unitary random matrices.

For the global correlation functions in the interior of the ellipse, we could only present partial 
results, based on the Chebyshev polynomials of the first, second and third kind. 

It would be very interesting to find the global asymptotic formulas for all families of Coulomb gases presented here, 
perhaps taking a closer look at the Riemann mapping theorem. A further direction of investigation is a comparison 
with the local correlations of both the standard elliptic Ginibre ensemble, and that with a hard constraint imposed. 
A popular tool in comparison with data is the number variance. Although it follows from the kernel it remains to be seen, 
if it could be further simplified in the various limits we have taken.

\section*{Acknowledgements}

The work of TN was partially supported by the Japan Society for the Promotion of Science (KAKENHI 25400397). 
GA \& MK acknowledge support by the German research council through CRC1283: "Taming uncertainty and profiting 
from randomness and low regularity in analysis, stochastics and their applications". GA also thanks the Niels 
Bohr International Academy for hospitality where part of this work was done. IP thanks support by the grant 
DAAD-CONICYT/Becas Chile, 2016/91609937. We thank Yacin Ameur for insightful discussions about the hard edge 
scaling limit.

\appendix
\section{Two families of asymmetric Jacobi polynomials}\label{AJacobi}
\setcounter{equation}{0}
\renewcommand{\theequation}{A.\arabic{equation}}

In this appendix we study the weak non-Hermiticity limit for two further families of planar orthogonal polynomials derived in \cite{ANPV}. 
Let us give a reason for the existence of these classes in addition to (\ref{orthogonality}). As we saw in the transformations 
(\ref{map1}) and ({\ref{map2}) of the Gegenbauer polynomials - which are usually expressed in terms of 
the Jacobi polynomials with symmetric indices \eqref{CP} - 
these can also be 
mapped to the Jacobi polynomials $P^{\left(a+\frac12,\pm \frac12\right)}_n(z)$. By using these mappings, 
one can see that the resulting Jacobi polynomials are also orthogonal on the 
same ellipse, but  with different weights. In this appendix, we evaluate the asymptotic 
behaviour of the Coulomb gas associated to those weights.

\subsection{The Jacobi polynomials $P^{\left(a+\frac12,\frac12\right)}_n(z)$}\label{Jacobi+1/2}

It is shown in \cite{ANPV} that the Jacobi polynomials $P^{\left(a+\frac12,\frac12\right)}_n(z)$ 
satisfy the orthogonality relation
\begin{eqnarray}
& & \int_E d^2z\ w_+(z)
P^{\left(a+\frac12,\frac12\right)}_m(z) P^{\left(a+\frac12,\frac12\right)}_n({\bar z}) 
\nonumber \\ 
& = & 4 \sqrt{\frac{1 - \tau}{2 \tau}} \frac{\Gamma(n + \frac32)^2 \Gamma(a + 1)^2}{(2 n + a + 2) 
\Gamma(n + a + 2)^2} C^{(a + 1)}_{2 n + 1}\left(\sqrt{\frac{1 + \tau}{2 \tau}}\right) \delta_{mn},
\end{eqnarray}
where $a > -1$, $E$ is the elliptic domain (\ref{ellipse}), and $C^{(a + 1)}_n(z)$ are 
the Gegenbauer polynomials (\ref{gegenbauer}). 
The 
one-particle weight function $w(z)$ in (\ref{pdf}) defining this type of Coulomb gas takes the form
\begin{equation}
\label{weight+}
w_+(z) =  ( 1 - \mu(z))^a,
\end{equation}
and 
\begin{equation}
\label{muz}
\mu(z) = \frac{2\tau}{1 - \tau} \left( \sqrt{\frac{ 1 + \tau}{2 \tau }} \sqrt{(1 + x)^2 + y^2} 
- 1 - x  \right),
\end{equation}
with $z = x + i y$. This weight function is different from (\ref{weight}),  except in the case $a = 0$, when the indices of the Jacobi polynomials again become symmetric. Note that the monic orthogonal polynomials 
$M_n(z) = z^n + \cdots
$ are given by \cite{NIST}
\begin{equation}
M_n(z) = 2^n n! \frac{\Gamma(n + a + 2)}{\Gamma(2 n + a + 2)} P^{\left(a +\frac12,\frac12\right)}_n(z).
\end{equation}
We obtain the kernel  $K_N(z_1,z_2)$ in (\ref{detkernel}) as
\begin{eqnarray}
\label{pluskernel}
K_N(z_1,z_2) &=& \frac{1}{4} (1 - \mu(z_1))^{a/2} (1 - \mu({\bar z}_2))^{a/2} \sqrt{\frac{2 \tau}{1 - \tau}}  \frac{1}{\Gamma(a + 1)^2} \nonumber \\ 
& & \times \sum_{n=0}^{N-1} \frac{(2 n + a + 2) \Gamma(n + a + 2)^2}{\Gamma(n + \frac32)^2 C^{(a + 1)}_{2 n + 1}\left( \sqrt{(1 + \tau)/(2 \tau)}  \right)}  
P^{\left(a + \frac12,\frac12\right)}_n(z_1) P^{\left(a + \frac12,\frac12\right)}_n({\bar z}_2).\nonumber \\ 
\end{eqnarray}
In the following, we will evaluate the asymptotic forms of this kernel in the weak non-Hermiticity limit at the edges, that is 
around the foci of the ellipse $z = +1$ and $z = -1$. Because in Section \ref{sec:weaklim} we have seen that the bulk limit 
can be recovered from the edge limit, we will first derive the latter. However, due to the indices of the Jacobi polynomials 
now being non-symmetric, we expect the limits at the endpoints $\pm1$ to be different, because of the lack of parity 
symmetry, cf. \eqref{Psym}. 

\par
\medskip
\noindent
(1) \underline{Edge limit at the focus $z = +1$:}
\par
\medskip
\noindent
In order to magnify this region, we recall the weak non-Hermiticity limit \eqref{weaklimit}
\begin{equation}
\label{taus}
\frac{1}{\tau} = 1 + \frac{s^2}{2 N^2},
\end{equation}
and the rescaling \eqref{Zedge} around the right focus $+1$:
\begin{equation}
\label{zjZj}
z_j = 1 - \frac{Z_j}{2 N^2}, \ \ \ j=1,2.
\end{equation}
We will take the double scaling limit $N \rightarrow \infty$ and $\tau\to1$ such that the positive number $s$ and 
complex numbers $Z_j = X_j + i Y_j$ are kept fixed. In this scaling limit the function inside the weight \eqref{weight+} 
gets mapped to
\begin{equation}
\label{mu+limit}
1 - \mu\left(1 - \frac{Z}{2 N^2} \right) 
\sim \frac{1}{4 N^2} \left( \frac{s^2}{4} + X - \frac{Y^2}{s^2} \right),
\end{equation}
from which we can read off the domain of our scaling variables, being  in the parabolic domain (\ref{para}). 
Here $Z = X + i Y$ is kept fixed. In analogy to \eqref{normlimit} we have 
\begin{equation}
\label{ci}
C^{(a + 1)}_{2 n + 1}\left(\sqrt{\frac{1 + \tau}{2 \tau}}\right) \sim N^{2 a + 1}
\frac{\Gamma(a + \frac32)}{\Gamma(2 a + 2)} \left( \frac{s}{8 c} \right)^{-a - \frac12} I_{a + \frac12}(c s),
\end{equation}
with the ratio $c = n/N$ being kept fixed. 
Using  (\ref{Jacobiasymptotic}), we can readily find the asymptotic for the polynomials
\begin{equation}
P^{\left(a + \frac12,\frac12\right)}_n\left( 1 - \frac{Z}{2 N^2} \right) 
\sim 
N^{a + \frac12} \left(\frac{\sqrt{Z}}{2} \right)^{- a - \frac12} J_{a + \frac12} \left(c \sqrt{Z}\right).
\end{equation}
Putting these asymptotic formulas together with the identities \eqref{tauid} into (\ref{pluskernel}), and replacing the sum by an integral, 
we obtain exactly the same 
asymptotic formula (\ref{edgekernel}) for 
$K_{\rm Edge}(Z_1,Z_2) = \lim_{N \rightarrow \infty} K_N(z_1,z_2) /(4 N^4)$.
This fact indicates the universality of this kernel. 

The Hermitian and strongly non-Hermitian limit as well as the bulk limit then follow as in Subsection \ref{weak-edge}.

\par
\medskip
\noindent
(2) \underline{Edge limit at the focus $z = -1$:}
\par
\medskip
\noindent
Next, we use the scaling in the weak non-Hermiticity limit (\ref{taus}) and magnify the region around the left focus $z = -1$ in the same way as in \eqref{zjZj}:
\begin{equation}
\label{zj-Zj}
 z_j = -1 + \frac{Z_j}{2 N^2}, \ \ \ j=1,2,
\end{equation}
with $s > 0$ and $Z_j = X_j + i Y_j$ fixed in the limit $N \rightarrow \infty$. 
It is straightforward to derive 
the asymptotic form of the weight function
\begin{equation}
\label{mu-limit}
1 - \mu\left( - 1 + \frac{Z}{2 N^2} \right) 
 \sim 1 - \frac{2}{s^2} \left( \sqrt{X^2 + Y^2} - X \right). 
\end{equation}
Here $Z = X + i Y$ is kept fixed. 
For this factor to be non-negative it can be seen that the points 
$(X_j, Y_j)$ have to lie inside the parabolic domain (\ref{para}). 
For the asymptotic form of the Jacobi polynomials with non-symmetric indices we have 
\begin{equation}
\label{Pasympt+}
P^{\left(a + \frac12,\frac12\right)}_n\left( -1 + \frac{Z}{2 N^2} \right) 
=(-1)^nP^{\left(\frac12,a + \frac12\right)}_n\left( 1 - \frac{Z}{2 N^2} \right) 
\sim 
(-1)^n N^{\frac12} \left(\frac{\sqrt{Z}}{2} \right)^{- \frac12} J_{\frac12} \left(c \sqrt{Z}\right),
\end{equation}
in the limit $N \rightarrow \infty$, after using \eqref{Psym} and \eqref{Jacobiasymptotic}.
These asymptotic formulas together with (\ref{ci}) 
are put into the kernel (\ref{pluskernel}) and yield
\begin{eqnarray}
\label{Kedge+1/2}
K_{\rm Edge}(Z_1,Z_2) &=& 
 \lim_{N \rightarrow \infty} \frac{1}{4N^4} K_N(z_1,z_2) \nonumber \\ & = & 
\frac{ (s/2)^{a - \frac{1}{2}} }{4\sqrt{\pi} \Gamma(a + 1)} 
\left( 1 - \frac{2}{s^2} \left( |Z_1| - X_1 \right) \right)^{a/2}
\left( 1 - \frac{2}{s^2} \left( |Z_2| - X_2 \right) \right)^{a/2}
 \nonumber \\ 
& & \times  \left( \sqrt{ Z_1 {\bar Z_2}} \right)^{- \frac{1}{2}} 
\int_0^1 dc \ \frac{c^{a + \frac{3}{2}}}{I_{a +  \frac{1}{2}}(c s)} 
J_{\frac{1}{2}}\left(c \sqrt{Z_1}\right) 
J_{\frac{1}{2}}\left(c \sqrt{\bar{Z_2}}\right).
 \nonumber \\ 
& = &  \frac{(s/2)^{a - \frac{1}{2}}  }{2\pi^{3/2} \Gamma(a + 1)} 
\left( 1 - \frac{2}{s^2} \left( |Z_1| - X_1 \right) \right)^{a/2}
\left( 1 - \frac{2}{s^2} \left( |Z_2| - X_2 \right) \right)^{a/2}\nonumber \\ 
& & \times \frac{1}{\sqrt{ Z_1 {\bar Z_2}}}
\int_0^1 dc \ \frac{c^{a + \frac{1}{2}}}{I_{a +  \frac{1}{2}}(c s)} \sin\left(c \sqrt{Z_1}\right) \sin\left(c \sqrt{\bar{Z_2}}\right) .
\end{eqnarray}
In the last step the $J$-Bessel functions are expressed in terms of sine,  using (\ref{J1/2}).  
For $a\neq0$ this edge kernel is clearly different from the one obtained for the Gegenbauer polynomials in \eqref{edgekernel} in Subsection \ref{weak-edge}.  While the local asymptotic form of the Jacobi polynomials around this focal point yields $J_{\frac12}$ (represented by means of the sine function), the influence of the edge is obviously still present through the dependence of the other factors on $a$.

In the Hermitian limit $s \rightarrow 0$,  the coordinates $(X_j,Y_j)$ are confined to 
the domain satisfying $X_j \geq 0$ and $Y_j  = 0$, as we saw already in the Subsection \ref{weak-edge}.
Using (\ref{ssmall}) and normalising the area as in \eqref{besselkernel}, we find the asymptotic formula
\begin{eqnarray}
& & 
\lim_{s \rightarrow 0}  \left. \frac{s\pi}{2(a+1)B} K_{\rm Edge}(Z_1,Z_2) 
\right|_{X_{1,2} \geq 0, Y_{1,2} = 0} 
\nonumber \\ & = & \frac{1}{4} \left( X_1 X_2\right)^{- \frac{1}{4}}
\int_0^1 dc \ c \ J_{ \frac{1}{2}}\left(c \sqrt{X_1} \right) 
J_{\frac{1}{2}}\left(c \sqrt{X_2} \right). 
\end{eqnarray}
It agrees with the Bessel-kernel of the Jacobi ensemble (\ref{besselkernel}) at $a = 0$. 

\par
In the strong non-Hermiticity limit $s \rightarrow \infty$ we use the scaling variables ${\tilde X}_j$ and 
${\tilde Y}_j$ defined in (\ref{tildescale}), together with the asymptotic relation
\begin{equation}
\label{zxx}
1 - \frac{2}{s^2} \left(\left| Z_j \right| - X_j \right) \sim \frac{2}{s} {\tilde X}_j, \ \ \ s \rightarrow \infty, 
\end{equation}
and (\ref{jcos}). The resulting limit  $ \lim_{s \rightarrow \infty} (s^2/4) K_{\rm Edge}(Z_1,Z_2)$ 
exactly reproduces the formula (\ref{sinfinity}). 

\par
The bulk limit $h \rightarrow \infty$, 
with the scaling variables ${\hat z}_j = {\hat x}_j + i {\hat y}_j$ 
defined as in \eqref{kappa} by $Z_j = \kappa h - 2 \sqrt{h} {\hat z}_j$ ($\kappa > 0$),  
can be evaluated by means of the relation
\begin{equation}
\label{zxy}
1 - \frac{2}{s^2} \left(\left| Z_j \right| - X_j \right) \sim 1 - \frac{4}{\kappa s^2} {\hat y}_j^2, \ \ \ h\rightarrow \infty\ ,
\end{equation}
and (\ref{jcos}).   As a result we obtain exactly the same formula (\ref{bulklimit}) 
for the asymptotic kernel 
$K_{\rm Bulk}({\hat z}_1,{\hat z}_2) = \lim_{h \rightarrow \infty} 
4 h K_{\rm Edge}(Z_1,Z_2)$. From this, we again conjecture that the bulk scaling limit has 
a similar form, when we zoom into any point $x_0 \in (-1,1)$. Thus all three limits of the kernel 
\eqref{Kedge+1/2} lead back to the classes we have already found in Section \ref{sec:weaklim}.

\par
\medskip
\noindent
\subsection{The Jacobi polynomials $P^{\left(a+\frac12,-\frac12\right)}_n(z)$}\label{Jacobi-1/2}

The Jacobi polynomials $P^{\left(a+\frac12,-\frac12\right)}_n(z)$ satisfy the orthogonality 
relation \cite{ANPV}
\begin{eqnarray}
& & \int_E d^2z \ 
w_-(z)
P^{\left(a+\frac12,-\frac12\right)}_m(z) P^{\left(a+\frac12,-\frac12\right)}_n({\bar z}) 
\nonumber \\ 
& = & 2 \sqrt{\frac{1 - \tau}{2 \tau}} \frac{\Gamma(n + \frac12)^2 \Gamma(a + 1)^2}{(2 n + a + 1) 
\Gamma(n + a + 1)^2} C^{(a + 1)}_{2 n}\left(\sqrt{\frac{1 + \tau}{2 \tau}}\right) \delta_{mn}, 
\end{eqnarray}
where $a > -1$, $E$ is the elliptic domain (\ref{ellipse}), and $C^{(a + 1)}_n(z)$ are
the Gegenbauer polynomials  (\ref{gegenbauer}). Moreover, $w(z)$ in (\ref{pdf}) takes 
the form
\begin{equation}
w_-(z) =  \frac{( 1 - \mu(z))^a}{|1 + z|},
\end{equation}
with $\mu(z)$ defined in (\ref{muz}). Notice that also for $a=0$ the polynomials and weight are 
different from those in Section \ref{corr}.

The monic orthogonal polynomials 
$M_n(z) = z^n + \cdots$ are given by \cite{NIST}
\begin{equation}
M_n(z) = 2^n n! \frac{\Gamma(n + a + 1)}{\Gamma(2 n + a + 1)} P^{\left(a +\frac12,-\frac12\right)}_n(z),
\end{equation}
and for the kernel function $K_N(z_1,z_2)$ in (\ref{detkernel}) we obtain from the above
\begin{eqnarray}
\label{minuskernel}
K_N(z_1,z_2) &=& \frac{(1 - \mu(z_1))^{a/2} (1 - \mu({\bar z}_2))^{a/2} }{2 |1 + z_1|^{1/2} |1 + z_2|^{1/2}} \sqrt{\frac{2 \tau}{1 - \tau}}  \frac{1}{\Gamma(a + 1)^2} \nonumber \\ 
& & \times \sum_{n=0}^{N-1} \frac{(2 n + a + 1) \Gamma(n + a + 1)^2}{\Gamma(n + \frac12)^2\ C^{(a + 1)}_{2 n}\left( \sqrt{(1 + \tau)/(2 \tau)}  \right)}
P^{\left(a + \frac12,-\frac12\right)}_n(z_1) P^{\left(a + \frac12,-\frac12\right)}_n({\bar z}_2).
\nonumber\\
\end{eqnarray}
As in the previous subsection we will first determine the weak non-Hermiticity limit at the edges. 
\par
\medskip
\noindent
(1) \underline{Edge limit at the focus $z = +1$}
\par
\medskip
\noindent
In the vicinity of the focus $+1$, we can again utilise the scalings (\ref{taus}) and 
(\ref{zjZj}), finding the same domain (\ref{para}) as before.  From (\ref{Jacobiasymptotic}), we find
\begin{equation}
P^{\left(a + \frac12,-\frac12\right)}_n\left( 1 - \frac{Z}{2 N^2} \right) 
\sim 
N^{a + \frac12} \left(\frac{\sqrt{Z}}{2} \right)^{- a - \frac12} J_{a + \frac12} \left(c \sqrt{Z}\right),
\end{equation}
in the limit $N \rightarrow \infty$. It agrees with \eqref{Pasympt+} because of its independence of the second index of the Jacobi polynomials.

We put this together with (\ref{ci}) - which does not change to leading order under the shift $2 n + 1 \mapsto 2 n$ -  and (\ref{mu+limit}) into (\ref{minuskernel}), and again 
find exactly the same asymptotic formula (\ref{edgekernel}) for $K_{\rm Edge}(Z_1,Z_2) 
=  \lim_{N \rightarrow \infty} K_N(z_1,z_2) /(4N^4)$. After the analysis of the previous subsection this universality is not unexpected.
The corresponding limits to Hermiticity, strong non-Hermiticity and the bulk thus follow alike.

\par
\medskip
\noindent
(2) \underline{Edge limit at the focus $z = -1$}
\par
\medskip
\noindent
Finally we use the scalings (\ref{taus}) and (\ref{zj-Zj}) to 
study the asymptotic behaviour of the kernel in the vicinity 
of $z = -1$. As in the previous subsection the coordinates $(X_j, Y_j)$ are in the domain (\ref{para}). For the asymptotic behaviour we now find
\begin{equation}
P^{\left(a + \frac12,-\frac12 \right)}_n\left( -1 + \frac{Z}{2 N^2} \right) 
\sim 
(-1)^n N^{-1/2} \left(\frac{\sqrt{Z}}{2} \right)^{1/2} J_{-1/2} \left(c \sqrt{Z}\right),
\end{equation}
in the limit $N \rightarrow \infty$, due to \eqref{Psym} and \eqref{Jacobiasymptotic}.
 This formula (\ref{ci}) being also true for shifted index $2 n + 1 \mapsto 2 n$, 
and (\ref{mu-limit}) are put into the kernel (\ref{minuskernel}). The result is
\begin{eqnarray}
\label{Kedge-1/2}
K_{\rm Edge}(Z_1,Z_2) &=& 
 \lim_{N \rightarrow \infty} \frac{1}{4N^4} K_N(z_1,z_2) \nonumber \\ & = & 
\frac{(s/2)^{a - \frac{1}{2}}  }{4\sqrt{\pi} \Gamma(a + 1)} 
\left( 1 - \frac{2}{s^2} \left( |Z_1| - X_1 \right) \right)^{a/2}  
\left( 1 - \frac{2}{s^2} \left( |Z_2| - X_2 \right) \right)^{a/2} 
 \nonumber \\ 
& & \times  \left( \frac{\sqrt{ Z_1 {\bar Z_2}}}{|Z_1 Z_2|} \right)^{ \frac{1}{2}} 
\int_0^1 dc \ \frac{c^{a + \frac{3}{2}}}{I_{a +  \frac{1}{2}}(c s)} 
J_{-\frac{1}{2}}\left(c \sqrt{Z_1}\right) 
J_{-\frac{1}{2}}\left(c \sqrt{\bar{Z_2}}\right)
 \nonumber \\ 
& = &  \frac{(s/2)^{a - \frac{1}{2}}  }{2\pi^{3/2} \Gamma(a + 1)} 
\left( 1 - \frac{2}{s^2} \left( |Z_1| - X_1 \right) \right)^{a/2}
\left( 1 - \frac{2}{s^2} \left( |Z_2| - X_2 \right) \right)^{a/2} 
 \nonumber \\ 
& & \times |Z_1 Z_2|^{-1/2}
\int_0^1 dc \ \frac{c^{a + \frac{1}{2}}}{I_{a +  \frac{1}{2}}(c s)} 
\cos\left(c \sqrt{Z_1}\right) 
\cos\left(c \sqrt{\bar{Z_2}}\right) .
\end{eqnarray}
In the last step we used (\ref{J-1/2}), expressing the $J$-Bessel functions through cosine.
Once again this edge kernel is different from that in \eqref{edgekernel} in Subsection \ref{weak-edge}, with the influence of the edge clearly visible through the dependence on $a$.

\par
In the Hermitian limit  $s \rightarrow 0$, 
we again put $(X_j,Y_j)$ in the domain satisfying $X_j \geq 0$ and $Y_j  = 0$. As before (\ref{ssmall}) leads to
\begin{eqnarray}
&& \lim_{s \rightarrow 0}  \left. \frac{s\pi}{2(a+1)B} K_{\rm Edge}(Z_1,Z_2) 
\right|_{X_{1,2} \geq 0, Y_{1,2} = 0} 
\nonumber \\ & = & \frac{1}{4} \left( X_1 X_2\right)^{-\frac{1}{4}}
\int_0^1 dc \ c \ J_{- \frac{1}{2}}\left(c \sqrt{X_1} \right) 
J_{-\frac{1}{2}}\left(c \sqrt{X_2} \right), 
\end{eqnarray}
which agrees with (\ref{besselkernel}) continued to $a = -1$,
\par
In the strong non-Hermiticity limit $s \rightarrow \infty$ we use the scalings (\ref{tildescale}) 
and the asymptotic relations (\ref{zxx}) and (\ref{jcos}).  It follows that 
$ \lim_{s \rightarrow \infty} (s^2/4) K_{\rm Edge}(Z_1,Z_2)$ is identical to 
the result in (\ref{sinfinity}). 

The bulk limit $h \rightarrow \infty$ with the scaling (\ref{kappa}) 
can be treated along the same line as in the previous subsection, by using (\ref{zxy}) and (\ref{jcos}).   We 
find exactly the same formula (\ref{bulklimit}) 
for $K_{\rm Bulk}({\hat z}_1,{\hat z}_2) = \lim_{h \rightarrow \infty} 
4 h K_{\rm Edge}(Z_1,Z_2)$.  We again conjecture that a similar bulk asymptotic form  holds 
for this model. Also for these polynomials all three limits lead back to known results.


\section{Correlations for Chebyshev polynomials of 1st kind}\label{BChebyshev}
\setcounter{equation}{0}
\renewcommand{\theequation}{B.\arabic{equation}}

In this appendix we will derive the limiting microscopic and global kernel for
the Chebyshev polynomials of the first kind $T_n(z)$. The orthogonality relation was previously known \cite{MH}, cf. \cite[Corollary 4.4]{ANPV}, but it does not follow directly from that of the 
Gegenbauer polynomials presented in Section \ref{corr}:
\begin{eqnarray}
\int_E d^2 z\  w_I(z) 
T_m(z) T_n({\bar z})  
&=& 
\left\{ \begin{array}{ll} \displaystyle \frac{\pi}{2 n} \sqrt{\frac{1 - \tau}{2 \tau}} C^{(1)}_{2 n - 1}\left(
\sqrt{\frac{1 + \tau}{2 \tau} } \right) \delta_{mn},&  
m>0, n \geq 0, 
\\ 
\displaystyle 
2 \pi \log v, & m=n=0,
\end{array} \right. 
\nonumber \\
\end{eqnarray}
on the ellipse (\ref{ellipse}), 
where
\begin{equation}
w_I(z) = \frac{1}{|1 - z^2|},
\end{equation}
and
 \begin{equation}
\label{vdef}
 v  = \frac{\sqrt{1+\tau}+\sqrt{1-\tau}}{\sqrt{2\tau}},
 \ \ \ 
 0 < \tau < 1\ .
 \end{equation}
The weight function $w(z)$ of the corresponding Coulomb gas model (\ref{pdf})  is 
given by $w_I(z)$, with singularities at the foci $\pm 1$. The Chebyshev 
polynomials of the 
first kind are
\begin{equation}
T_n(z) = \sqrt{\pi} \frac{\Gamma(n + 1)}{\Gamma(n + \frac12)} P^{\left(-\frac12,-\frac12\right)}_n(z), 
\end{equation}
in terms of the Jacobi polynomials, which are again symmetric. Note that the corresponding 
monic orthogonal polynomials are $M_n(z) = 2^{-n + 1} T_n(z)$ for $n \geq 1$, 
and $M_0(z)=T_0(z) = 1$. We find the kernel  in (\ref{kernel}) with $w(z) = w_I(z)$ as
\begin{eqnarray}
\label{firstkernel}
K_N(z_1,z_2) &=& \frac{1}{\pi} \frac{1}{\sqrt{|1 - z_1^2| |1 - {\bar z}_2^2|}} 
\nonumber \\ &&  \times  \left[ \sqrt{\frac{2 \tau}{1 - \tau}}  
\sum_{n=1}^{N-1} \frac{2 n}{
C^{(1)}_{2 n - 1}\left(\sqrt{(1 + \tau)/( 2 \tau)}\right)} T_n(z_1) T_n({\bar z}_2) 
+ \frac{1}{2 \log v} \right]. \nonumber \\ 
\end{eqnarray}

\subsection{Local edge scaling limit}
In order to evaluate the asymptotic behaviour of this kernel around 
the focus $z=1$, we again adopt the scalings (\ref{taus}) and (\ref{zjZj}). 
Because the polynomials have parity we only need to analyse one of the foci.
It can readily be seen from (\ref{Jacobiasymptotic}) that
\begin{equation}
T_n\left( 1 - \frac{Z}{2 N^2} \right) \sim \sqrt{\pi} c^{1/2} \left( 
\frac{\sqrt{Z}}{2} \right)^{1/2} J_{-\frac{1}{2}}\left(c \sqrt{Z} \right),
\end{equation}
in the limit $N \rightarrow \infty$ with $c = n/N$ fixed.  We use this formula, 
 (\ref{ci}) valid to leading order at shifted index $2 n + 1 \mapsto 2 n - 1$, and the expansion 
 \begin{equation}
 v^2  \sim 1 + \frac{s}{N},  \ \ \ N \rightarrow \infty,
 \end{equation}
that follows from \eqref{tauid}.
Inserting them into (\ref{firstkernel}) we obtain
 \begin{eqnarray}
K_{\rm Edge}(Z_1,Z_2) &=& 
 \lim_{N \rightarrow \infty} \frac{1}{4N^4} K_N(z_1,z_2) \nonumber \\ & = & 
 \frac14
\sqrt{\frac{2}{s \pi }}  \left( \frac{\sqrt{ Z_1 {\bar Z_2}}}{|Z_1 Z_2|} \right)^{ \frac{1}{2}} 
\int_0^1 dc \ \frac{c^{\frac{3}{2}}}{I_{\frac{1}{2}}(c s)} 
J_{-\frac{1}{2}}\left(c \sqrt{Z_1}\right) 
J_{-\frac{1}{2}}\left(c \sqrt{\bar{Z_2}}\right), \quad\quad
 \end{eqnarray}
which is identical to (\ref{Kedge-1/2}) with $a = 0$. Consequently, this kernel is also universal, 
and the corresponding Hermitian, strongly non-Hermitian and bulk limits follow as discussed in 
Subsection \ref{Jacobi-1/2}.

\subsection{Global correlations}
In order to derive a global asymptotic formula for the kernel, we use the relations \cite{MH}
\begin{equation}
T_n\left(\frac{z}{\sqrt{2 \tau}}\right) = \frac{1}{2} \left( \omega^n + \frac{1}{\omega^n} \right), 
\ \ \ \frac{z}{\sqrt{2 \tau} }= \frac{1}{2} \left(\omega + \frac{1}{\omega} \right),
\end{equation}
and
\begin{equation}
\sqrt{ \frac{1 - \tau}{2 \tau}} C^{(1)}_{2 n - 1}\left(\sqrt{\frac{1 + \tau}{2 \tau}} \right) 
= \sqrt{ \frac{1 - \tau}{2 \tau}} U_{2 n - 1}\left(\sqrt{\frac{1 + \tau}{2 \tau}} \right) =  \frac{1}{2} \left( v^{2 n} - \frac{1}{v^{2 n}} \right)
\end{equation}
(from \eqref{wv} together with \eqref{Urel}), in (\ref{firstkernel}) to find
\begin{eqnarray}
K_N\left(\frac{z_1}{\sqrt{2 \tau}}, \frac{z_2}{\sqrt{2 \tau}}\right) &=& \frac{1}{\pi} 
\left| 1 - \frac{z_1^2}{2 \tau} \right|^{-1/2} \left| 1 - \frac{{\bar z}_2^2}{2 \tau} \right|^{-1/2} 
\nonumber \\ 
& & \times \left[ \sum_{n=1}^{N-1} \frac{n}{v^{2 n} - v^{-2 n}}  
\left( \omega_1^n + \frac{1}{\omega_1^n} \right) \left( {\bar \omega}_2^n + \frac{1}{{\bar \omega}_2^n} 
\right) + \frac{1}{2 \log v} \right].\quad\quad
\end{eqnarray}
Here, we define $z_1/\sqrt{2 \tau} = (\omega_1 + \omega_1^{-1})/2$  and 
$z_2/\sqrt{2 \tau} = (\omega_2 + \omega_2^{-1})/2$ .  As $z_1/\sqrt{2 \tau} $ and 
$z_2/\sqrt{2 \tau}$ are in the interior of the ellipse (\ref{ellipse}), 
$1 \leq |\omega_1| < v$ and $1 \leq |\omega_2| < v$. Now, in order to take 
the limit $N \rightarrow \infty$, up to a pre-factor,  we can use the same argument as in Section \ref{weight1}. 
The result is
\begin{eqnarray}
K_{\rm global}(z_1,z_2)
&=& \lim_{N \rightarrow \infty} \frac{1}{2 \tau} K_N\left(\frac{z_1}{\sqrt{2 \tau}},\frac{z_2}{\sqrt{2 \tau}}\right) \nonumber \\ 
& = & \frac{1}{2 \pi \tau} \left| 1 - \frac{z_1^2}{2 \tau} \right|^{-1/2} 
\left| 1 - \frac{{\bar z}_2^2}{2 \tau} \right|^{-1/2} \nonumber \\ && \times  
\left[ \sum_{j=0}^{\infty} \left(
\frac{\eta_j \omega_1 {\bar \omega}_2}{(1 - (\eta_j \omega_1 {\bar \omega}_2))^2}
+ \frac{\eta_j \omega_1/{\bar \omega}_2}{(1 - (\eta_j \omega_1/{\bar \omega}_2))^2} \right. \right.
\nonumber \\ & & \left. \left.
+ \frac{\eta_j {\bar \omega}_2/\omega_1}{(1 - (\eta_j {\bar \omega}_2/\omega_1))^2}
+ \frac{\eta_j/(\omega_1 {\bar \omega}_2)}{(1 - (\eta_j/(\omega_1 {\bar \omega}_2)))^2} 
\right)  + \frac{1}{2 \log v} \right],
\end{eqnarray}
with $\eta_j = 1/v^{2(1 + 2 j)}$.  

Moreover, we can take the radially symmetric 
limit $\tau \rightarrow 0$ ($v \rightarrow \infty$) and introducing the same scaling arguments as in \eqref{ozeta} to obtain
\begin{equation}
\label{1stgk}
\lim_{\tau \rightarrow 0} K_{\rm global}(z_1,z_2)
= \frac{1}{\pi |z_1||z_2|} \frac{z_1 {\bar z}_2}{(1 - z_1 {\bar z_2})^2},
\end{equation}
when $0 < |z_j| < 1$, $j = 1,2$. In the domain of validity this kernel is equivalent 
to the one in \eqref{Krot}.


\section{Correlations for Chebyshev polynomials of 3rd kind}\label{CChebyshev}
\setcounter{equation}{0}
\renewcommand{\theequation}{C.\arabic{equation}}

A similar procedure to that in Appendix B can be applied to the Coulomb gas 
model (\ref{pdf}), with the weight function $w(z)$ given by
\begin{equation}
w_{III}(z) = \frac{1}{|1 + z| } \ \ {\rm or} \ \ w_{IV}(z) = \frac{1}{|1 - z| } .
\end{equation}
These models correspond to the Chebyshev polynomials of the third and fourth kind, 
respectively. It should be noted that the model with $w_{III}(z)$ is a special case $a = 0$ 
of the model studied in Subsection \ref{Jacobi-1/2}. Moreover the model with $w_{IV}(z)$ 
is an image under the mapping $z \rightarrow -z$ of the model with $w_{III}(z)$.  
Therefore, in the following, we only  treat the global asymptotic formulas for the 
model with $w_{III}(z)$. The corresponding Chebyshev polynomials of 
the third kind 
 \begin{equation}
 V_n(z) = \frac{2 n + 1}{P^{\left(\frac12,-\frac12\right)}_n(1)} P^{\left(\frac12,-\frac12\right)}_n(z),
 \end{equation}
 satisfy the orthogonality relation\cite{MH}
 \begin{equation}
\int_E d^2z \ w_{III}(z) V_m(z) V_n({\bar z}) 
= \frac{2 \pi}{2 n + 1}  \sqrt{\frac{1 - \tau}{2 \tau}} C^{(1)}_{2 n}\left( \sqrt{\frac{1 + \tau}{2 \tau} } \right) \delta_{mn},
\end{equation}
with $m,n = 0,1,2,\cdots$.  The corresponding monic orthogonal polynomials are 
 $M_n(z) = 2^{-n} V_n(z)$. We use the relations \cite{MH}
\begin{equation}
V_n\left(\frac{z}{\sqrt{2 \tau}} \right) = \frac{\omega^{n + \frac{1}{2}} + 
\omega^{- n - \frac{1}{2}}}{w^{\frac{1}{2}} + w^{-\frac{1}{2}}}, \ \ \ 
\frac{z}{\sqrt{2 \tau} }= \frac{1}{2} \left(\omega + \frac{1}{\omega} \right),
\end{equation}
and
\begin{equation}
\sqrt{ \frac{1 - \tau}{2 \tau}} C^{(1)}_{2 n}\left(\sqrt{\frac{1 + \tau}{2 \tau}} \right) 
= \sqrt{ \frac{1 - \tau}{2 \tau}} U_{2 n}\left(\sqrt{\frac{1 + \tau}{2 \tau}} \right) =  
\frac{1}{2} \left( v^{2 n + 1} - \frac{1}{v^{2 n + 1}} \right),
\end{equation}
to obtain the kernel function in (\ref{detkernel}) with $w(z) = w_{III}(z)$. Here, $v$ is defined in 
(\ref{vdef}). The result is
\begin{eqnarray}
& & K_N\left(\frac{z_1}{\sqrt{2 \tau}}, \frac{z_2}{\sqrt{2 \tau}}\right) = \frac{1}{\pi} 
\left| 1 +  \frac{z_1}{\sqrt{2 \tau}} \right|^{-1/2} \left| 1 +  \frac{{\bar z}_2}{\sqrt{2 \tau}} \right|^{-1/2} 
\nonumber \\ 
& & \times \sum_{n=0}^{N-1} \frac{2 n + 1}{v^{2 n+1} - v^{-2 n-1}}  
\frac{\left( \omega_1^{n+\frac{1}{2}} + \omega_1^{-n-\frac{1}{2}}  \right) \left( {\bar \omega}_2^{n+\frac{1}{2}} +{\bar \omega}_2^{-n-\frac{1}{2}} \right) }
{ \left( \omega_1^{\frac{1}{2}} + \omega_1^{-\frac{1}{2}} \right) \left( {\bar \omega}_2^{\frac{1}{2}} 
+ {\bar \omega}_2^{-\frac{1}{2}} \right)},
\end{eqnarray}
where $z_1/\sqrt{2 \tau} = (\omega_1 + (1/\omega_1))/2$  and $z_2/\sqrt{2 \tau} = (\omega_2 + (1/\omega_2))/2$,  and again
 $1 \leq |\omega_1| < v$ and $1 \leq |\omega_2| < v$.  As before we use an argument similar to that employed in Section \ref{weight1} and Appendix \ref{BChebyshev}, and find
\begin{eqnarray}
K_{\rm global}(z_1,z_2)
& = & \lim_{N \rightarrow \infty} \frac{1}{2 \tau} K_N\left(\frac{z_1}{\sqrt{2 \tau}},\frac{z_2}{\sqrt{2 \tau}}\right) \nonumber \\ 
& = & \frac{1}{2 \pi \tau} \left| 1 + \frac{z_1}{\sqrt{2 \tau}} \right|^{-1/2} 
\left| 1 + \frac{{\bar z}_2}{\sqrt{2 \tau}} \right|^{-1/2} 
\frac{1}{ \left( \omega_1^{\frac{1}{2}} + \omega_1^{-\frac{1}{2}} \right) \left( {\bar \omega}_2^{\frac{1}{2}} 
+ {\bar \omega}_2^{-\frac{1}{2}} \right)} \nonumber \\ & & \times \sum_{j=0}^{\infty} \left[
\frac{(\eta_j \omega_1 {\bar \omega}_2)^{1/2} (1 + (\eta_j \omega_1 {\bar \omega}_2))}{
(1 - (\eta_j \omega_1 {\bar \omega}_2))^2}
+ \frac{(\eta_j \omega_1/{\bar \omega}_2)^{1/2} (1 + (\eta_j \omega_1/{\bar \omega}_2))}{
(1 - (\eta_j \omega_1/{\bar \omega}_2))^2} \right. 
\nonumber \\ & & \left. 
+ \frac{(\eta_j {\bar \omega}_2/\omega_1)^{1/2} (1 + (\eta_j {\bar \omega}_2/\omega_1))}{
(1 - (\eta_j {\bar \omega}_2/\omega_1))^2}
+ \frac{(\eta_j/(\omega_1 {\bar \omega}_2))^{1/2} (1 + (\eta_j/(\omega_1 {\bar \omega}_2)))}{
(1 - (\eta_j/(\omega_1 {\bar \omega}_2)))^2} 
\right], \nonumber \\
\end{eqnarray}
with $\eta_j = 1/v^{2(1 + 2 j)}$.  We can moreover take the radially symmetric 
limit $\tau \rightarrow 0$ ($v \rightarrow \infty$) and obtain
\begin{equation}
\lim_{\tau \rightarrow 0} K_{\rm global}(z_1,z_2)
= \frac{1}{2 \pi \sqrt{|z_1||z_2|}} \frac{1 + z_1 {\bar z}_2}{(1 - z_1 {\bar z_2})^2},
\end{equation}
when $0 < |z_j| < 1$ ($j = 1,2$). It differs from (\ref{1stgk}) and (\ref{Krot}).


\end{document}